\documentclass[12pt,fleqn,a4paper]{article}
\usepackage{epsfig}
\usepackage[dvips]{color}
\begin{document}
\title{Exotic quark effects on the Higgs sector of the USSM at the LHC}
\author{S. W. Ham$^{(1,2)}$ and S. K. Oh$^{(3)}$
\\
\\
{\it $^{(1)}$ Department of Physics, KAIST, Daejon 305-701, Korea}
\\
{\it $^{(2)}$ School of Physics, KIAS, Seoul 130-722, Korea}
\\
{\it $^{(3)}$ Department of Physics, Konkuk University, Seoul 143-701, Korea}
\\
\\
}
\date{}
\maketitle
\begin{abstract}
The Higgs sector of the $U(1)$-extended supersymmetric model is studied with great detail.
We calculate the masses of the Higgs bosons at the one-loop level.
We also calculate at the one-loop level the gluon-involving processes for the productions and decays of the
scalar Higgs bosons of the model at the energy of the CERN Large Hadron Collider (LHC), where the radiative
corrections due to the loops of top, bottom, and exotic quarks and their scalar partners are taken into account.
We find that the exotic quark and exotic scalar quarks in the model may manifest themselves at the LHC,
since the production of the heaviest scalar Higgs boson via gluon fusion processes is mediated virtually by
the loops of exotic quark and exotic scalar quarks, for a reasonable parameter set of the model.
\end{abstract}


\section{Introduction}

The search for the Higgs boson is one of the prominent goals of the future high-energy collider experiments.
The massive fermions and gauge bosons in the standard model (SM) acquire their masses
through the mediation of a single scalar Higgs boson.
The mass of this SM Higgs boson receives quadratic divergences at the one-loop level due
to the radiative corrections from other SM particle loops.
The simplest way to solve this problem is to introduce the supersymmetry (SUSY) to the SM.
The SUSY partners to the SM particles help to eliminate the quadratic divergences to the Higgs boson mass,
by the cancellation between each SM particle loop and its superparticle loop, at the one-loop level.
However, if the SUSY is a broken symmetry in nature, the cancelation is not complete and thus
the Higgs boson mass receives non-zero quantum corrections at the one-loop level [1-4].

The most economical SUSY model is the minimal supersymmetric standard model (MSSM), where
two Higgs doublets are introduced in order to cancel the gauge anomaly.
They give masses independently to down-type fermions and up-type fermions.
Since studies of the MSSM has revealed a few theoretical problems such as the $\mu$ problem [5],
various alternative SUSY models have been introduced.

The $U(1)$-extended supersymmetric standard model (USSM) is one of the non-minimal
supersymmetric models which does not suffer the MSSM $\mu$ problem.
The USSM is characterized by the extra $U(1)'$ gauge symmetry at the TeV scale [6-23].
Thus, there is an extra neutral gauge boson with a mass of a TeV scale.
Also, the USSM has a larger Higgs sector, which consists of a neutral Higgs singlet besides two Higgs doublets.
In the USSM, the quantity corresponding to the $\mu$ parameter of the MSSM is dynamically
generated by the vacuum expectation value of the Higgs singlet.

Another characteristic of the USSM is that each SM generation is extended by an extra pair
of $SU(2)$ singlet quarks, $D_L^l$ and ${\bar D}^l_R$, with electric charges $-1/3$ and $+1/3$, respectively.
The presence of these exotic quarks is interesting from the phenomenological point of view.
In particular, scenarios with a light exotic quark in the USSM have extensively been investigated [18-22].
A recent study have shown that the exotic quarks may contribute by large amount to the CP mixing
between the heaviest scalar and pseudoscalar Higgs bosons [18].
Since the exotic quarks couple directly to the neutral Higgs singlet, the heaviest Higgs boson mass
at the one-loop level might be affected by the exotic quark loops.
Also, the exotic quark loops contribute to the gluon fusion process of the Higgs production
at the Large Hadron Collider (LHC).
These effects of the exotic quarks have been studied elsewhere [22].
While some approximations have been made in Ref. [22],
the calculations in this article are carried out in a rigorous way.

In this article, we continue to study the Higgs potential in the USSM at the one-loop level.
In particular, we are interested in the Higgs productions and decays at the LHC.
We take into account the radiative corrections due to the quarks of the third generation,
exotic quarks, and their superpartners at the leading order.
We calculate the production cross sections of the scalar Higgs bosons via the gluon fusion process.
It is found that the production of the heaviest scalar Higgs boson might be significantly enhanced for
a parameter region by the exotic quark and scalar quarks contributions.

We also study the dominant decay modes of the scalar Higgs bosons.
Among them, we calculate the decay processes of the scalar Higgs bosons into gluon pairs,
which are mediated by the loops of the quarks, exotic quarks, and their superpartners.
The gluonic decay mode might be important for the Higgs search of the USSM at the LHC, since it
may exhibit the role of the exotic quark and exotic scalar quarks.
In particular, we find that the decay of the heaviest scalar Higgs boson into a pair of gluons depends
virtually only on the loops of the exotic quark and exotic scalar quarks.

The article is organized as follows: In the next section, the Higgs potential of the USSM is described.
Then, we calculate the masses of the pseudoscalar and the three neutral scalar Higgs bosons at
the one-loop level. These one-loop masses are taken as input for the Higgs production calculations
in Section 4 and the Higgs decay calculations in Section 5.
Numerical analysis is presented in Section 6, and conclusions in the last section.

\section{Higgs Potential}

There are two Higgs doublet superfields ${\cal H}_1$ and ${\cal H}_2$, and a Higgs singlet superfield
${\cal S}$ in the USSM.
The gauge symmetry of the USSM is $SU (2) \times U(1) \times U(1)'$.
For the Higgs superpotential, we consider only the quarks of the third generation, and one family of
the exotic quarks, $D_L$ and ${\bar D}_R$.
Thus, the superpotential of the USSM may be written as
\begin{equation}
    W \approx h_t Q^T \epsilon {\cal H}_2 t_R^c - h_b Q^T \epsilon {\cal H}_1 b_R^c + h_k {\cal S} D_L {\bar D}_R
    + \lambda {\cal S} {\cal H}_1^T \epsilon {\cal H}_2  \ ,
\end{equation}
where $h_t$, $h_b$, and $h_k$ are the Yukawa couplings of the top, bottom, and exotic quarks, respectively,
$Q$ is the left-handed quark doublet superfield of the third generation,
$t_R^c$ and $b_R^c$ are respectively the charge conjugate superfields of the right-handed top quark
and the right-handed bottom quark,
$\lambda$ is the dimensionless coupling constant,
and $\epsilon$ is a $2 \times 2$ antisymmetric matrix defined as $\epsilon_{12} = 1$.

The Higgs sector of the USSM consists of two Higgs doublets, $H_1 = (H_1^0, H^-)$ and $H_2 = (H^+, H_2^0)$,
and one Higgs singlet $S$.
The tree-level Higgs potential, $V^0$, is obtained by collecting the $F$-terms and the $D$-terms of the
above superpotential, and adding physically allowed soft terms.
Thus, it may be written as
\begin{equation}
V^0 = V_F + V_D + V_{\rm S} \ ,
\end{equation}
with
\begin{eqnarray}
V_F & = & |\lambda|^2 [(|H_1|^2 + |H_2|^2) |S|^2 + |H_1^T \epsilon H_2|^2]  \ , \cr
V_D & = & {g_2^2 \over 8} (H_1^{\dagger} \vec\sigma H_1 + H_2^{\dagger} \vec\sigma H_2)^2
+ {g_1^2 \over 8} (|H_1|^2 - |H_2|^2)^2 \cr
& &\mbox{}+ {g^{'2}_1 \over 2} ( {\tilde Q}_1 |H_1|^2 + {\tilde Q}_2 |H_2|^2 + {\tilde Q}_3 |S|^2)^2 \ , \cr
V_{\rm S} & = & m_1^2 |H_1|^2 + m_2^2 |H_2|^2 + m_3^2 |S|^2
- [\lambda A_{\lambda} H_1^T \epsilon H_2 S + {\rm H.c.}] \ ,
\end{eqnarray}
where $g_2$, $g_1$, and $g'_1$ are respectively the gauge coupling coefficients of $SU(2)$, $U(1)$, and $U(1)'$,
$\vec\sigma$ are Pauli matrices, $A_{\lambda}$ is the trilinear soft SUSY breaking parameter with mass dimension,
$m_i$ ($i$ = 1,2,3) are soft SUSY breaking masses, and ${\tilde Q}_1$, ${\tilde Q}_2$,
and ${\tilde Q}_3$ are respectively the effective $U(1)'$ hypercharges of $H_1$, $H_2$, and $S$.
We do not consider in this article the CP violation in the USSM, which may take place in the mixing between the
scalar and pseudoscalar Higgs bosons.
Thus, the parameters of the Higgs potential as well as the vacuum expectation values of the
neutral Higgs fields are taken to be real in following procedure.

The effective $U(1)'$ hypercharges of the Higgs fields satisfy the identity relation of
$\sum_{i=1}^3 {\tilde Q}_i = 0$ in order to ensure the $U(1)'$ gauge invariance.
The soft SUSY breaking masses $m_i$ may be eliminated by using the three minimum conditions for
the Higgs potential that define the vacuum.
After electroweak symmetry breaking, the neutral components of Higgs fields will develop vacuum expectation
values as $v_1 = \langle H_1^0 \rangle$, $v_2 = \langle H_2^0 \rangle$, and $s = \langle S \rangle$.
We introduce a free parameter, $\tan \beta = v_2/v_1$.
The value of $v = \sqrt{v_1^2+v_2^2}$ is fixed by electroweak data as 175 GeV.
In terms of the vacuum expectation values, the masses of top, bottom, and exotic quarks are
respectively given as $m_t = h_t v_2$,  $m_b = h_b v_1$, and $m_k = h_k s$
after electroweak symmetry breaking.

The tree-level $2 \times 2$ mass matrices for the scalar partners to top, bottom, and exotic quarks
in the USSM may be written as
\begin{equation}
M^{\tilde q} =
\left(\begin{array}{cc}
M^{\tilde q}_{11} & M^{\tilde q}_{12}  \\
M^{{\tilde q}  *}_{12} & M^{\tilde q}_{22}  \
\end{array} \right)  \ ,
\end{equation}
where ${\tilde q} = {\tilde t}, {\tilde b}, {\tilde k}$.
For simplicity, we rewrite $M^{\tilde q}_{ij}$ as $M_{ij}^{\tilde t}$, $M_{ij}^{\tilde b}$, and $M_{ij}^{\tilde k}$, respectively,
for the scalar partners to top, bottom, and exotic quarks.
The matrix elements are obtained from the tree-level Higgs potential in the on-shell Lagrangian after the elimination
of the auxiliary fields as follows:
\begin{eqnarray}
M_{11}^{\tilde t} & = & m_Q^2 + h_t^2 |H_2^0|^2
+ \left ({g_2^2 \over 4} - {g_1^2 \over 12} \right ) (|H_1^0|^2 - |H_2^0|^2) \cr
& &\mbox{} + {g^{'2}_1 \over 4} {\tilde Q}_Q ({\tilde Q}_1 |H_1^0|^2 + {\tilde Q}_2 |H_2^0|^2 + {\tilde Q}_3 |S|^2) \ , \cr
M_{22}^{\tilde t} & = & m_U^2 + h_t^2 |H_2^2|^2 + {g_1^2 \over 3} (|H_1^0|^2 - |H_2^0|^2) \cr
& &\mbox{} + {g^{'2}_1 \over 4} {\tilde Q}_U ({\tilde Q}_1 |H_1^0|^2 + {\tilde Q}_2 |H_2^0|^2 + {\tilde Q}_3 |S|^2) \ , \cr
M_{12}^{\tilde t} & = & h_t (\lambda H_1^{0 *} S^* - A_t H_2^0) \ , \cr
M_{11}^{\tilde b} & = & m_Q^2 + h_b^2 |H_1^0|^2 - \left ({g_2^2 \over 4} + {g_1^2 \over 12} \right )(|H_1^0|^2 - |H_2^0|^2)   \cr
& &\mbox{} + {g^{'2}_1 \over 4} {\tilde Q}_Q ({\tilde Q}_1 |H_1^0|^2 + {\tilde Q}_2 |H_2^0|^2 + {\tilde Q}_3 |S|^2) \ , \cr
M_{22}^{\tilde b} & = & m_D^2 + h_b^2 |H_1^0|^2 - {g_1^2 \over 6} (|H_1^0|^2 - |H_2^0|^2) \cr
& &\mbox{} + {g^{'2}_1 \over 4} {\tilde Q}_D ({\tilde Q}_1 |H_1^0|^2 + {\tilde Q}_2 |H_2^0|^2 + {\tilde Q}_3 |S|^2) \ , \cr
M_{12}^{\tilde b} & = & h_b (\lambda H_2^{0 *} S^* - A_b H_1^0) \ , \cr
M_{11}^{\tilde k} & = & m_K^2 + h_k^2 |S|^2 + {g_1^2 \over 6} (|H_1^0|^2 - |H_2^0|^2)\cr
& &\mbox{} + {g^{'2}_1 \over 4} {\tilde Q}_K ({\tilde Q}_1 |H_1^0|^2 + {\tilde Q}_2 |H_2^0|^2 + {\tilde Q}_3 |S|^2) \ , \cr
M_{22}^{\tilde k} & = & m_{\bar K}^2 + h_k^2 |S|^2 - {g_1^2 \over 6} (|H_1^0|^2 - |H_2^0|^2)\cr
& &\mbox{} + {g^{'2}_1 \over 4} {\tilde Q}_{\bar K} ({\tilde Q}_1 |H_1^0|^2 + {\tilde Q}_2 |H_2^0|^2 + {\tilde Q}_3 |S|^2) \ , \cr
M_{12}^{\tilde k} & = & h_k (\lambda H_1^{0 *} H_2^{0 *} - A_k S) \ ,
\end{eqnarray}
where $m_Q$, $m_U$, $m_D$, $m_K$ and $m_{\bar K}$ are the soft SUSY breaking masses for the scalar quarks,
$A_t$, $A_b$, $A_k$ are the trilinear soft SUSY breaking parameters for them,
and
${\tilde Q}_Q = - {\tilde Q}_1/3$, ${\tilde Q}_U = ( {\tilde Q}_1 - 2 {\tilde Q}_2 )/3$,
${\tilde Q}_D = ( {\tilde Q}_1 + 2 {\tilde Q}_2 )/3$, ${\tilde Q}_K = {\tilde Q}_2$, and
${\tilde Q}_{\tilde K} = {\tilde Q}_1$.
These relations for the effective $U(1)'$ hypercharges are dictated by the cancellation of anomalies.
Note that the contributions from $V_D$ are included in the above mass matrices for the scalar quarks.
In Ref. [22], they have been neglected.
Thus,  if we set $g_1 = g_2 = g'_1 = 0$,
the above expressions will reduce to the corresponding results in Ref. [22].

Now, the squared masses of the scalar partners to top, bottom, and exotic quarks are respectively given by
the eigenvalues of $M^{\tilde q}$ (${\tilde q} = {\tilde t}, {\tilde b}, {\tilde k}$), by diagonalizing them.
We obtain the squared masses of the scalar quarks as
\begin{eqnarray}
m_{{\tilde t}_1, {\tilde t}_2}^2 & = & {m_Q^2 + m_U^2 \over 2} + m_t^2 + {m_Z^2 \over 4} \cos 2 \beta
+ G_1^t v^2 \cos^2 \beta  + G_2^t v^2 \sin^2 \beta + G_3^t {\tilde Q}_3 s^2 \cr
& &\mbox{} \mp \sqrt{X_t} \ , \cr
m_{{\tilde b}_1, {\tilde b}_2}^2 & = & {m_Q^2 + m_D^2 \over 2} + m_b^2 - {m_Z^2 \over 4} \cos 2 \beta
+ G_1^b v^2 \cos^2 \beta  + G_2^b v^2 \sin^2 \beta + G_3^b s^2 \cr
& &\mbox{} \mp \sqrt{X_b} \ , \cr
m_{{\tilde k}_1, {\tilde k}_2}^2 & = & {m_K^2 + m_{\bar K}^2 \over 2} + m_k^2
+ G_1^k v^2 \cos^2 \beta  + G_2^k v^2 \sin^2 \beta + G_3^k s^2 \mp \sqrt{X_k}    \ ,
\end{eqnarray}
where the contributions of the scalar quark mixing are given by
\begin{eqnarray}
X_t & = & \left [ {m_Q^2 - m_U^2 \over 2} + \left ( {2 m_W^2 \over 3} - {5 m_Z^2 \over 12} \right ) \cos 2 \beta
+ B_1^t v^2 \cos^2 \beta  + B_2^t \sin^2 \beta + B_3^t s^2     \right ]^2 \cr
& &\mbox{} + m_t^2 (\lambda s \cot \beta - A_t)^2    \   ,   \cr
X_b & = & \left [ {m_Q^2 - m_D^2 \over 2} + \left ( {m_Z^2 \over 12} - {m_W^2 \over 3} \right ) \cos 2 \beta
+ B_1^b v^2 \cos^2 \beta  + B_2^b v^2 \sin^2 \beta + B_3^b s^2     \right ]^2 \cr
& &\mbox{} + m_b^2 (\lambda s \tan \beta - A_b)^2    \   , \cr
X_k & = & \left [ {m_K^2 - m_{\bar K}^2 \over 2} + \left ( {m_Z^2 \over 3} - {m_W^2 \over 3} \right ) \cos 2 \beta
+ B_1^k v^2 \cos^2 \beta  + B_2^k v^2 \sin^2 \beta + B_3^k s^2     \right ]^2 \cr
& &\mbox{} + m_k^2 (\lambda v^2 \sin \beta \cos \beta /s - A_k)^2    \  ,
\end{eqnarray}
with
\begin{eqnarray}
G_i^t & = & {g^{'2}_1 \over 8} ({\tilde Q}_Q + {\tilde Q}_U) {\tilde Q}_i \ , \cr
G_i^b & = & {g^{'2}_1 \over 8} ({\tilde Q}_Q + {\tilde Q}_D) {\tilde Q}_i \ , \cr
G_i^k & = & {g^{'2}_1 \over 8} ({\tilde Q}_K + {\tilde Q}_{\bar K}) {\tilde Q}_i \ , \cr
B_i^t & = & {g^{'2}_1 \over 8} ({\tilde Q}_Q - {\tilde Q}_U) {\tilde Q}_i \ , \cr
B_i^b & = & {g^{'2}_1 \over 8} ({\tilde Q}_Q - {\tilde Q}_D) {\tilde Q}_i \ , \cr
B_i^k & = & {g^{'2}_1 \over 8} ({\tilde Q}_K - {\tilde Q}_{\bar K}) {\tilde Q}_i \ .
\end{eqnarray}

\section{Higgs Masses at the One-Loop Level}

At the one-loop level, quarks and their scalar partners contribute to the radiative corrections through loops.
The contributions from the loops of top quark and top scalar quarks are most dominant for
a wide region in the parameter space.
Also, the radiative corrections due to the loops of bottom quark and bottom scalar quarks might be
phenomenologically significant, especially for very large $\tan \beta$, at the low energy scale.
The contributions from the loops of the exotic quark and its scalar partners in the USSM are not well studied.
We have suggested elsewhere that their contributions are worthwhile studying from a phenomenological point of view
because the exotic quark couples directly to the Higgs singlet $S$ in the USSM [22].
Thus, in this article, we consider the loops of top, bottom, exotic quark and their scalar partners in our calculations
at the one-loop level.

The Higgs potential at the one-loop level, $V^1$, is given as
\[
    V^1 = V^0 + V_{1,eff}
\]
where the radiative corrections, $V_{1,eff}$, is obtained by effective potential method [24].
Explicitly, it is given by
\begin{equation}
V_{1,eff}  = \sum_{j} {n_j {\cal M}_j^4 \over 64 \pi^2}
\left [
\log {{\cal M}_j^2 \over \Lambda^2} - {3 \over 2}
\right ]  \ ,
\end{equation}
where $\Lambda$ is the renormalization scale in the modified minimal subtraction scheme,
and $n_q = -12$ is the quark degree of freedom and $n_{{\tilde q}_i} = 6$ ($i=1,2$) are
the degrees of freedom for scalar quarks, determined from their color, charge, and spin factors.

After the electroweak symmetry breaking, ten real degrees of freedom from the Higgs sector of the USSM
are reduced to six physical Higgs particles, namely, a pair of charged Higgs bosons, one pseudoscalar Higgs boson,
and three scalar Higgs bosons.
The squared mass of the neutral pseudoscalar Higgs boson at the one-loop level may be written as
\begin{equation}
m_A^2 = m_{A^0}^2 + m_{A^t}^2 + m_{A^b}^2  + m_{A^k}^2  \ ,
\end{equation}
where $m_{A^0}$ is the tree-level mass, given as
\[
m_{A^0}^2 = {2 \lambda v A_{\lambda} \over \sin 2 \alpha} \ ,
\]
and $m_{A^t}$, $m_{A^b}$, and $m_{A^k}$ account respectively for the radiative corrections from the
top, bottom, and the exotic quark sectors:
\begin{eqnarray}
m_{A^t}^2 & = &\mbox{} - {3 m_t^2 \lambda A_t \over 8 \pi^2 v \sin^2 \beta \sin 2 \alpha}
f(m_{{\tilde t}_1}^2, \ m_{{\tilde t}_2}^2)   \ , \cr
m_{A^b}^2 & = &\mbox{} - {3 m_b^2 \lambda A_b \over 8 \pi^2 v \cos^2 \beta \sin 2 \alpha}
f(m_{{\tilde b}_1}^2, \ m_{{\tilde b}_2}^2)   \ , \cr
m_{A^k}^2 & = &\mbox{} - {3 m_k^2 \lambda A_k v \over 8 \pi^2 s^2 \sin 2 \alpha}
f(m_{{\tilde k}_1}^2, \ m_{{\tilde k}_2}^2)   \ ,
\end{eqnarray}
where $\alpha$ is a mixing angle between the electroweak scale and the $U(1)'$ symmetry breaking scale, defined as
\begin{equation}
\tan \alpha = {v \over 2 s} \sin 2 \beta \ ,
\end{equation}
and the dimensionless function $f(m_1^2, \ m_2^2)$ is defined as
\begin{equation}
 f(m_1^2, \ m_2^2) = {1 \over (m_2^2 - m_1^2)} \left[  m_1^2 \log {m_1^2 \over \Lambda^2} - m_2^2
\log {m_2^2 \over \Lambda^2} \right] + 1 \ .
\end{equation}

The three neutral scalar Higgs bosons, $S_i$ ($i=1,2,3$), of the USSM are given
by the eigenvectors of a $3 \times 3$ symmetric mass matrix,
and their squared masses, $m_{S_i}$ ($i = 1,2,3$), are given by the corresponding eigenvalues.
These neutral scalar Higgs bosons are sorted such that $m_{S_1} < m_{S_2} < m_{S_2}$.

At the one-loop level, the mass matrix for the three scalar Higgs bosons, may
be decomposed as
\[
    M_{ij} = M_{ij}^0 + M_{ij}^t + M_{ij}^b + M_{ij}^k
\]
where $M_{ij}^0$ is the matrix elements at the tree level, obtained from $V^0$, given explicitly as
\begin{eqnarray}
M_{11}^0 & = & m_Z^2 \cos^2 \beta + 2 g'^2_1 {\tilde Q}_1^2 v^2 \cos^2 \beta
+ m_{A^0}^2 \sin^2 \beta \cos^2 \alpha  \ ,  \cr
M_{22}^0 & = & m_Z^2 \sin^2 \beta + 2 g'^2_1 {\tilde Q}_2^2 v^2 \sin^2 \beta
+ m_{A^0}^2 \cos^2 \beta \cos^2 \alpha \ ,  \cr
M_{33}^0 & = & 2 g'^2_1 {\tilde Q}_3^2 s^2 + m_{A^0}^2 \sin^2 \alpha \ , \cr
M_{12}^0 & = & g'^2_1 {\tilde Q}_1 {\tilde Q}_2 v^2 \sin 2 \beta + (\lambda^2 v^2 - m_Z^2/2) \sin 2 \beta
- m_{A^0}^2 \cos \beta \sin \beta \cos^2 \alpha \ ,  \cr
M_{13}^0 & = & 2 g'^2_1 {\tilde Q}_1 {\tilde Q}_3 v s \cos \beta + 2 \lambda^2 v s \cos \beta
- m_{A^0}^2 \sin \beta \cos \alpha \sin \alpha \ , \cr
M_{23}^0 & = & 2 g'^2_1 {\tilde Q}_2 {\tilde Q}_3 v s \sin \beta + 2 \lambda^2 v s \sin \beta
- m_{A^0}^2 \cos \beta \cos \alpha \sin \alpha  \ ,
\end{eqnarray}
and $M_{ij}^t$, $M_{ij}^b$, and $M_{ij}^k$ are respectively the radiative contributions from the top quark sector,
the bottom quark sector, and the exotic quark sector.
They are obtained from $V_{1,eff}$ as
\begin{eqnarray}
M_{ij}^q & = & {3 \over 32 \pi^2 v^2} W_i^q W_j^q
{g(m_{{\tilde q}_1}^2, m_{{\tilde q}_2}^2) \over (m_{{\tilde q}_2}^2 - m_{{\tilde q}_1}^2)^2}
+ {3 \over 32 \pi^2 v^2} A_i^q A_j^q \log \left ( {m_{{\tilde q}_1}^2 m_{{\tilde q}_2}^2 \over \Lambda^4 } \right ) \cr
& &\mbox{} + {3 \over 32 \pi^2 v^2} (W_i^q A_j^q + A_i^q W_j^q)
{ \log ( m_{{\tilde q}_2}^2/ m_{{\tilde q}_1}^2)  \over (m_{{\tilde q}_2}^2 - m_{{\tilde q}_1}^2)} + D_{ij}^q   \  ,
\end{eqnarray}
where  $q = t,b,k$ and
\begin{eqnarray}
A_1^t & = & (2 G_1^t v^2 + {m_Z^2 \over 2}) \cos \beta \ , \cr
A_2^t & = & {2 m_t^2 \over \sin \beta} + (2 G_2^t v^2 - {m_Z^2 \over 2}) \sin \beta  \ , \cr
A_3^t & = & 2 G_3^t v s   \ ,  \cr
A_1^b & = & {2 m_b^2 \over \cos \beta} + (2 G_1^b v^2 - {m_Z^2 \over 2}) \cos \beta \ , \cr
A_2^b & = &  (2 G_2^b v^2 + {m_Z^2 \over 2}) \sin \beta \ , \cr
A_3^b & = & 2 G_3^b v s   \ ,   \cr
A_1^k & = & 2 G_1^k v^2 \cos \beta \ , \cr
A_2^k & = & 2 G_2^k v^2 \sin \beta \ , \cr
A_3^k & = & {2 m_k^2 v \over s} + 2 G_3^k v s \ ,
\end{eqnarray}
\begin{eqnarray}
W_1^t & = & {2 m_t^2 \lambda s \Delta_t \over \sin \beta } + (2 B_1^t v^2 + {4 m_W^2 \over 3} - {5 m_Z^2 \over 6} ) \cos \beta \Delta_t^g  \ , \cr
W_2^t & = &\mbox{} - {2 m_t^2 A_t \Delta_t \over \sin \beta } - (-2 B_2^t v^2 + {4 m_W^2 \over 3} - {5 m_Z^2 \over 6} ) \sin \beta \Delta_t^g  \ , \cr
W_3^t & = & {2 m_t^2 \lambda v \Delta_t \over \tan \beta} +  2 B_3^t v s \Delta_t^g \ ,  \cr
W_1^b & = &\mbox{} - {2 m_b^2 A_b \Delta_b \over \cos \beta } + (2 B_1^b v^2 - {2 m_W^2 \over 3} + {m_Z^2 \over 6} ) \cos \beta \Delta_b^g  \ , \cr
W_2^b & = & {2 m_b^2 \lambda s \Delta_b \over \cos \beta } - (-2 B_2^b v^2 - {2 m_W^2 \over 3} + {m_Z^2 \over 6} ) \sin \beta \Delta_b^g  \ , \cr
W_3^b & = & 2 m_b^2 \lambda v \tan \beta \Delta_b +  2 B_3^b v s \Delta_b^g  \ , \cr
W_1^k & = & {2 m_k^2 \lambda v^2 \sin \beta \Delta_k \over s} + (2 B_1^k v^2 + {2 m_Z^2 \over 3} - {2 m_Z^2 \over 3} ) \cos \beta \Delta_k^g  \ , \cr
W_2^k & = & {2 m_k^2 \lambda v^2 \cos \beta \Delta_k \over s} - (-2 B_2^k v^2 + {2 m_Z^2 \over 3} - {2 m_Z^2 \over 3} ) \sin \beta \Delta_k^g  \ , \cr
W_3^k & = &\mbox{} - {2 A_k m_k^2 v \over s} \Delta_k + 2 B_3^k v s \Delta_k^g \ ,
\end{eqnarray}
\begin{eqnarray}
\Delta_t & = & \lambda s \cot \beta - A_t \ , \cr
\Delta_t^g & = & m_Q^2 - m_U^2 + ({4 \over 3} m_W^2 - {5 \over 6} m_Z^2) \cos 2 \beta
+ 2 B_1^t v^2 \cos^2 \beta + 2 B_2^t  v^2 \sin^2 \beta
+ 2 B_3^t s^2   \  ,  \cr
\Delta_b & = & \lambda s \tan \beta - A_b \ , \cr
\Delta_b^g & = & m_Q^2 - m_D^2 + ({1 \over 6} m_Z^2 - {2 \over 3} m_W^2) \cos 2 \beta
+ 2 B_1^b v^2 \cos^2 \beta + 2 B_2^b  v^2 \sin^2 \beta
+ 2 B_3^b s^2   \  ,                            \cr
\Delta_k & = & \lambda v \tan \alpha - A_k \ , \\
\Delta_k^g & = & m_K^2 - m_{\bar K}^2 + {2 \over 3} (m_Z^2 - m_W^2) \cos 2 \beta
+ 2 B_1^k v^2 \cos^2 \beta + 2 B_2^k v^2 \sin^2 \beta
+ 2 B_3^k s^2   \  ,  \nonumber
\end{eqnarray}
\begin{eqnarray}
D_{11}^t & = & m_{A^t}^2 \sin^2 \beta \cos^2 \alpha  - {3 \cos^2 \beta \over 16 \pi^2 v^2}
\left( 2 B_1^t v^2 + {4 m_W^2 \over 3} - {5 m_Z^2 \over 6} \right)^2
f(m_{{\tilde t}_1}^2, \ m_{{\tilde t}_2}^2) \ , \cr
D_{22}^t & = &  m_{A^t}^2 \cos^2 \beta \cos^2 \alpha
- {3 \sin^2 \beta \over 16 \pi^2 v^2}
\left( - 2 B_2^t v^2 + {4 m_W^2 \over 3} - {5 m_Z^2 \over 6} \right)^2
f(m_{{\tilde t}_1}^2, \ m_{{\tilde t}_2}^2)     \cr
& &\mbox{} - {3 m_t^4 \over 4 \pi^2 v^2 \sin^2 \beta} \log \left ({m_t^2 \over \Lambda^2} \right )  \ ,  \cr
D_{33}^t & = &  m_{A^t}^2 \sin^2 \alpha  - {(B_3^t s)^2 \over 4 \pi^2} f(m_{{\tilde t}_1}^2, \ m_{{\tilde t}_2}^2)  \ , \cr
D_{12}^t & = &\mbox{} - m_{A^t}^2 \cos \beta \sin \beta \cos^2 \alpha
+ {3 \sin 2 \beta \over 32 \pi^2 v^2}
\left (2 B_1^t v^2 + {4 m_W^2 \over 3} - {5 m_Z^2 \over 6} \right)  \cr
& &\mbox{} \times \left (- 2 B_2^t v^2 + {4 m_W^2 \over 3} - {5 m_Z^2 \over 6} \right)
f(m_{{\tilde t}_1}^2, \ m_{{\tilde t}_2}^2)  \ , \cr
D_{13}^t & = &\mbox{} - m_{A^t}^2 \sin \beta \cos \alpha \sin \alpha
- {3 m_t^2 \lambda^2 s \cos \beta \over 8 \pi^2 v \sin^2 \beta}
f(m_{{\tilde t}_1}^2, \ m_{{\tilde t}_2}^2) \cr
& &\mbox{} - {3 B_3^t s \cos \beta \over 8 \pi^2 v} \left (2 B_1^t v^2 + {4 m_W^2 \over 3} - {5 m_Z^2 \over 6} \right)
f(m_{{\tilde t}_1}^2, \ m_{{\tilde t}_2}^2)    \ , \cr
D_{23}^t & = &\mbox{} - m_{A^t}^2 \cos \beta \cos \alpha \sin \alpha   \cr
& &\mbox{} + {3 B_3^t s \sin \beta \over 8 \pi^2 v} \left (- 2 B_2^t v^2 + {4 m_W^2 \over 3} - {5 m_Z^2 \over 6} \right)
f(m_{{\tilde t}_1}^2, \ m_{{\tilde t}_2}^2)    \ ,
\end{eqnarray}
\begin{eqnarray}
D_{11}^b & = & m_{A^b}^2 \sin^2 \beta \cos^2 \alpha  - {3 \cos^2 \beta \over 16 \pi^2 v^2}
\left( 2 B_1^b v^2 - {2 m_W^2 \over 3} + {m_Z^2 \over 6} \right)^2
f(m_{{\tilde b}_1}^2, \ m_{{\tilde b}_2}^2)  \cr
& &\mbox{} - {3 m_b^4 \over 4 \pi^2 v^2 \cos^2 \beta} \log \left ({m_b^2 \over \Lambda^2} \right )  \ ,  \cr
D_{22}^b & = &  m_{A^b}^2 \cos^2 \beta \cos^2 \alpha
- {3 \sin^2 \beta \over 16 \pi^2 v^2}
\left( - 2 B_2^b v^2 - {2 m_W^2 \over 3} + {m_Z^2 \over 6} \right)^2
f(m_{{\tilde b}_1}^2, \ m_{{\tilde b}_2}^2)     \  , \cr
D_{33}^b & = &  m_{A^b}^2 \sin^2 \alpha  - {(B_3^b s)^2 \over 4 \pi^2} f(m_{{\tilde b}_1}^2, \ m_{{\tilde b}_2}^2)  \ , \cr
D_{12}^b & = &\mbox{} - m_{A^t}^2 \cos \beta \sin \beta \cos^2 \alpha
+ {3 \sin 2 \beta \over 32 \pi^2 v^2}
\left (2 B_1^b v^2 - {2 m_W^2 \over 3} + {m_Z^2 \over 6} \right)  \cr
& &\mbox{} \times \left (- 2 B_2^b v^2 - {2 m_W^2 \over 3} + {m_Z^2 \over 6} \right)
f(m_{{\tilde t}_1}^2, \ m_{{\tilde t}_2}^2)  \ , \cr
D_{13}^b & = &\mbox{} - m_{A^b}^2 \sin \beta \cos \alpha \sin \alpha \cr
& &\mbox{} - {3 B_3^b s \cos \beta \over 8 \pi^2 v} \left (2 B_1^b v^2 - {2 m_W^2 \over 3} + {m_Z^2 \over 6} \right)
f(m_{{\tilde b}_1}^2, \ m_{{\tilde b}_2}^2)    \ , \cr
D_{23}^b & = &\mbox{} - m_{A^b}^2 \cos \beta \cos \alpha \sin \alpha
- {3 m_b^2 \lambda^2 s \tan \beta \over 8 \pi^2 v \cos \beta}
f(m_{{\tilde b}_1}^2, \ m_{{\tilde b}_2}^2) \cr
& &\mbox{} + {3 B_3^b s \sin \beta \over 8 \pi^2 v} \left (- 2 B_2^b v^2 - {2 m_W^2 \over 3} + {m_Z^2 \over 6} \right)
f(m_{{\tilde b}_1}^2, \ m_{{\tilde b}_2}^2)    \ ,
\end{eqnarray}
\begin{eqnarray}
D_{11}^k & = & m_{A^k}^2 \sin^2 \beta \cos^2 \alpha  - {3 \cos^2 \beta \over 16 \pi^2 v^2}
\left( 2 B_1^k v^2 + {2 m_Z^2 \over 3} - {2 m_W^2 \over 3} \right)^2
f(m_{{\tilde k}_1}^2, \ m_{{\tilde k}_2}^2) \ , \cr
D_{22}^k & = &  m_{A^k}^2 \cos^2 \beta \cos^2 \alpha
- {3 \sin^2 \beta \over 16 \pi^2 v^2}
\left( - 2 B_2^k v^2 + {2 m_Z^2 \over 3} - {2 m_W^2 \over 3} \right)^2
f(m_{{\tilde k}_1}^2, \ m_{{\tilde k}_2}^2)    \  , \cr
D_{33}^k & = &  m_{A^k}^2 \sin^2 \alpha  - {(B_3^k s)^2 \over 4 \pi^2} f(m_{{\tilde k}_1}^2, \ m_{{\tilde k}_2}^2)
- {3 m_k^4 \over 4 \pi^2 s^2} \log \left ({m_k^2 \over \Lambda^2} \right )  \ ,  \cr
D_{12}^k & = &\mbox{} - m_{A^k}^2 \cos \beta \sin \beta \cos^2 \alpha
+ {3 \sin 2 \beta \over 32 \pi^2 v^2}
\left (2 B_1^k v^2 + {2 m_Z^2 \over 3} - {2 m_W^2 \over 3} \right)  \cr
& &\mbox{} \times \left (- 2 B_2^k v^2 + {2 m_Z^2 \over 3} - {2 m_W^2 \over 3} \right)
f(m_{{\tilde k}_1}^2, \ m_{{\tilde k}_2}^2)  \cr
& &\mbox{} - {3 m_k^2 \lambda^2 v^2 \sin 2 \beta \over 16 \pi^2 s^2}
f(m_{{\tilde k}_1}^2, \ m_{{\tilde k}_2}^2)    \  ,   \cr
D_{13}^k & = &\mbox{} - m_{A^k}^2 \sin \beta \cos \alpha \sin \alpha
- {3 m_k^2 \lambda^2 s \cos \beta \over 8 \pi^2 v \sin^2 \beta}
f(m_{{\tilde k}_1}^2, \ m_{{\tilde k}_2}^2) \cr
& &\mbox{} - {3 B_3^k s \cos \beta \over 8 \pi^2 v} \left (2 B_1^k v^2 + {2 m_Z^2 \over 3} - {2 m_W^2 \over 3} \right)
f(m_{{\tilde k}_1}^2, \ m_{{\tilde k}_2}^2)    \ , \cr
D_{23}^k & = &\mbox{} - m_{A^k}^2 \cos \beta \cos \alpha \sin \alpha   \cr
& &\mbox{} + {3 B_3^k s \sin \beta \over 8 \pi^2 v} \left ( - 2 B_2^k v^2 + {2 m_Z^2 \over 3} - {2 m_W^2 \over 3} \right)
f(m_{{\tilde k}_1}^2, \ m_{{\tilde k}_2}^2)    \ .
\end{eqnarray}

The above analytic expressions are obtained by employing some mathematical formulas.
We remark that, in a previous study,
we have neglected for the sake of simplicity the $D$-term contributions of the the scalar quarks for
the neutral Higgs boson masses [22].
One may obtain the results of Ref. [22] by taking $g_1 = g_2 = g'_1 =0$ in the above expressions.
Thus, the above expressions for the scalar Higgs boson masses at the one-loop level are more rigorous
and complete, and calculated for the first time in the USSM.

\section{Higgs Productions in $pp$ Collisions}

It is well noticed that the gluon fusion process, mediated by the top quark triangular loops, is the most dominant production
channel for the SM Higgs boson in $pp$ collisions at the energy of the LHC.
The gluon fusion process is also important for the productions of the scalar Higgs bosons of the USSM at the LHC, where
the loops of top quark and top scalar quarks, by far, would play the most dominant role,
for a large space of the relevant parameters of the USSM.
For very large $\tan \beta$, the loops of bottom quark and bottom scalar quarks would also yield considerable contributions.
Further, the loops of the exotic quark and exotic scalar quarks may also contribute as much
significantly as the loops of top quark sector or the bottom quark sector.
Thus, in our analysis, we consider all of them, namely, the contributions of top, bottom, and
 the exotic quark sectors to the productions of the USSM scalar Higgs bosons via the gluon fusion process.

Let us denote ${\hat \sigma}_i (gg \to S_i)$ as the parton-level cross section for the gluon-gluon
annihilation into $S_i$ ($i =1,2,3$) in the USSM.
It is obtained as [25,26,22]
\begin{equation}
{\hat \sigma}_i ({\hat s} ) = {\hat \sigma}_i (gg \to S_i) = {\alpha_s^2(m_Z) \over 256 \pi}
    \delta \left (1-{m_{S_i}^2 \over {\hat s}^2 }\right )
    (| A^S + A_D^S|^2) \ ,
\end{equation}
where  ${\hat s}$ is the square of the c.m. energy of two gluons, $\alpha_s (m_Z)$
is the strong coupling constant evaluated at the electroweak scale,
and $A^S$ represents the contributions from top and bottom quarks and their scalar partners
while $A_D^S$ accounts for the contributions from the exotic quark sector.
They are given in terms of form factors as
\begin{eqnarray}
A^S & = & \sum_{q=t,b} \left[ G_{S_iqq} A_q^S (\tau_q)
+ \sum_{j=1,2} G_{S_i {\tilde q}_j {\tilde q}_j}
{v^2 \over 2 m_{{\tilde q}_j}^2} A^S_{\tilde q} (\tau_{{\tilde q}_j})
 \right ] \ ,   \cr
A_D^S & = & \left[ G_{S_ikk} A_k^S (\tau_k)
+ \sum_{j=1,2} G_{S_i {\tilde k}_j {\tilde k}_j }
{v^2 \over m_{{\tilde k}_j}^2} A^S_{\tilde k} (\tau_{{\tilde k}_j})
 \right ] \ .
\end{eqnarray}
where $A^S_q (\tau_q)$ is the quark form factor ($q = t,b,k$),
$G_{S_i qq}$ is the coupling coefficient of $S_i$ to a pair of quarks ($q = t,b,k$),
$A^S_{\tilde q} (\tau_{\tilde q})$ is the form factor of the scalar quarks (${\tilde q} = {\tilde t}, {\tilde b}, {\tilde k}$),
and $G_{S_i {\tilde q}_j {\tilde q}_j}$ are the coupling coefficients of $S_i$ to the pairs of
scalar quarks ($j=1,2$ and ${\tilde q} = {\tilde t}, {\tilde b}, {\tilde k}$).

The quark form factor and the form factor for the scalar quarks are given respectively as
\begin{equation}
A^S_q (\tau_q) = \tau_q [1 + (1 - \tau_q) f(\tau_q) ] \ , \quad
A^S_{\tilde q} (\tau_{{\tilde q}_j}) = \tau_{{\tilde q}_j} [ \tau_{{\tilde q}_j} f(\tau_{{\tilde q}_j}) - 1]  \  ,
\end{equation}
where the scaled variables are defined as
\begin{equation}
\tau_q = {4 m^2_q \over m^2_{S_i} } \ , \quad
\tau_{{\tilde q}_j} = {4 m^2_{{\tilde q}_j} \over m^2_{S_i} }   \ ,
\end{equation}
and the function $f$ is defined as
\begin{equation}
f(\tau) = \left \{
\begin{array}{cl}
{\rm arcsin}^2(1/\sqrt{\tau})    & \qquad \tau \geq 1 \ , \cr
-{1 \over 4} \left[ \log \left( {1+\sqrt{1+ \tau} \over 1 - \sqrt{1- \tau}} \right) -i \pi \right]^2
& \qquad \tau < 1 \ .
\end{array} \right .
\end{equation}

Now, let us calculate the coupling coefficients, $G_{S_i qq}$ and $G_{S_i {\tilde q}_j {\tilde q}_j}$.
It is convenient to normalize the coupling coefficients of $S_i$ to a pair of quarks using the corresponding SM coupling coefficient.
Then, the normalized coupling coefficients for our case are obtained as
\begin{equation}
G_{S_i bb}  =  {O_{1i} \over \cos \beta} \ , \quad
G_{S_i tt}  =  {O_{2i} \over \sin \beta} \ ,
\end{equation}
where $O_{ij}$ are the elements of the $3 \times 3$ orthogonal matrix which diagonalizes the mass matrix
for the scalar Higgs bosons.
The coupling coefficient of $S_i$ to a pair of exotic quarks, normalized with the exotic quark mass $m_k$, is given as
\begin{equation}
G_{S_i kk}  =  {v \over s} O_{3i}     \ .
\end{equation}

The coupling coefficients of $S_i$ to a pair of scalar quarks are defined as
\begin{equation}
G_{S_i {\tilde q}_j {\tilde q}_k }
= \sum^3_{l =1} \left ( \Gamma^{S_l {\tilde q} {\tilde q} }   \right )_{\beta \gamma}
O_{l i} U_{\beta j}^{\tilde q} U_{\gamma k}^{\tilde q} \ ,
\end{equation}
where $\Gamma^{S_i {\tilde q} {\tilde q} }$ ($i = 1,2,3$ and ${\tilde q} = {\tilde t}, {\tilde b}, {\tilde k}$)
are the $2 \times 2$ matrix for the coupling coefficients of $S_i$ to a pair of the left- and right-handed
 components of weak eigenstates of scalar quarks,
$U^{\tilde q}$ is the $2 \times 2$ orthogonal matrix which diagonalizes the mass matrix for the scalar quarks,
and the subscript indices are $\beta, \gamma=L,R$, and $j,k=1,2$.
Note that $O_{ij}$ are also present in the above expression.

The explicit expressions for $\Gamma^{S_i {\tilde q} {\tilde q} }$ are as follows:
\begin{eqnarray}
\Gamma^{S_1 {\tilde t} {\tilde t} } & = & {m_t \over v^2 \sin \beta}
\left(\begin{array}{cc}
0 & \lambda s  \\
\lambda s & 0  \
\end{array} \right) \cr
& &\mbox{} + v \cos \beta
\left(\begin{array}{cc}
{\displaystyle  -{g_1^2 \over 6} + {g_2^2 \over 2} + { g^{'2}_1 \over 8} {\tilde Q}_Q {\tilde Q}_1 }& 0  \\
0 & {\displaystyle  {2 \over 3} g_1^2 + { g^{'2}_1 \over 8} {\tilde Q}_U {\tilde Q}_1 } \
\end{array} \right)  \ , \cr
\Gamma^{S_2 {\tilde t} {\tilde t} } & = & {m_t \over v^2 \sin \beta}
\left(\begin{array}{cc}
2 m_t & - A_t  \\
- A_t & 2 m_t  \
\end{array} \right)  \cr
& &\mbox{} + v \sin \beta
\left(\begin{array}{cc}
{\displaystyle  {g_1^2 \over 6} - {g_2^2 \over 2} + { g^{'2}_1 \over 8} {\tilde Q}_Q {\tilde Q}_2 }& 0  \\
0 & {\displaystyle  -{2 \over 3} g_1^2 + { g^{'2}_1 \over 8} {\tilde Q}_U {\tilde Q}_2 } \
\end{array} \right)  \ , \cr
\Gamma^{S_3 {\tilde t} {\tilde t} } & = & {m_t \over v \sin \beta}
\left(\begin{array}{cc}
0 & \lambda \cos \beta  \\
\lambda \cos \beta & 0  \
\end{array} \right)  \cr
& &\mbox{} + s
\left(\begin{array}{cc}
{\displaystyle  { g^{'2}_1 \over 8} {\tilde Q}_Q {\tilde Q}_3 }& 0  \\
0 & {\displaystyle  { g^{'2}_1 \over 8} {\tilde Q}_U {\tilde Q}_3 } \
\end{array} \right)  \ , \cr
\Gamma^{S_1 {\tilde b} {\tilde b} } & = & {m_b \over v^2 \cos \beta}
\left(\begin{array}{cc}
2 m_b & - A_b \\
- A_b & 2 m_b  \
\end{array} \right)   \cr
& &\mbox{} + v \cos \beta
\left(\begin{array}{cc}
{\displaystyle  -{g_1^2 \over 6} - {g_2^2 \over 2} + { g^{'2}_1 \over 8} {\tilde Q}_Q {\tilde Q}_1 }& 0  \\
0 & {\displaystyle  -{ g_1^2 \over 3} + { g^{'2}_1 \over 8} {\tilde Q}_D {\tilde Q}_1 } \
\end{array} \right)  \ , \cr
\Gamma^{S_2 {\tilde b} {\tilde b} } & = & {m_b \over v^2 \cos \beta}
\left(\begin{array}{cc}
0 & \lambda s  \\
\lambda s & 0  \
\end{array} \right)   \cr
& &\mbox{} + v \sin \beta
\left(\begin{array}{cc}
{\displaystyle  {g_1^2 \over 6} + {g_2^2 \over 2} + { g^{'2}_1 \over 8} {\tilde Q}_Q {\tilde Q}_2 }& 0  \\
0 & {\displaystyle  { g_1^2 \over 3} + { g^{'2}_1 \over 8} {\tilde Q}_D {\tilde Q}_2 } \
\end{array} \right)  \ , \cr
\Gamma^{S_3 {\tilde b} {\tilde b} } & = & {m_b \over v \cos \beta}
\left(\begin{array}{cc}
0 & \lambda \sin \beta  \\
\lambda \sin \beta & 0  \
\end{array} \right)   \cr
& &\mbox{} + s
\left(\begin{array}{cc}
{\displaystyle  { g^{'2}_1 \over 8} {\tilde Q}_Q {\tilde Q}_3 }& 0  \\
0 & {\displaystyle  { g^{'2}_1 \over 8} {\tilde Q}_D {\tilde Q}_3 } \
\end{array} \right)  \ , \cr
\Gamma^{S_1 {\tilde k} {\tilde k} } & = & {m_k \over s}
\left(\begin{array}{cc}
0 & \lambda \sin \beta  \\
\lambda \sin \beta & 0  \
\end{array} \right)   \cr
& &\mbox{} + v \cos \beta
\left(\begin{array}{cc}
{\displaystyle  {g_1^2 \over 3} + { g^{'2}_1 \over 8} {\tilde Q}_K {\tilde Q}_1 }& 0  \\
0 & {\displaystyle  -{g_1^2 \over 3} + { g^{'2}_1 \over 8} {\tilde Q}_{\bar K} {\tilde Q}_1 } \
\end{array} \right)  \ , \cr
\Gamma^{S_2 {\tilde k} {\tilde k} } & = & {m_k \over s}
\left(\begin{array}{cc}
0 & \lambda \cos \beta  \\
\lambda \cos \beta & 0  \
\end{array} \right)   \cr
& &\mbox{} + v \sin \beta
\left(\begin{array}{cc}
{\displaystyle  - {g_1^2 \over 3} + { g^{'2}_1 \over 8} {\tilde Q}_K {\tilde Q}_2 }& 0  \\
0 & {\displaystyle  {g_1^2 \over 3} + { g^{'2}_1 \over 8} {\tilde Q}_{\bar K} {\tilde Q}_2 } \
\end{array} \right)  \ , \cr
\Gamma^{S_3 {\tilde k} {\tilde k} } & = & {m_k \over v s}
\left(\begin{array}{cc}
2 m_k & - A_k  \\
- A_k & 2 m_k  \
\end{array} \right)   \cr
& &\mbox{} + s
\left(\begin{array}{cc}
{\displaystyle  { g^{'2}_1 \over 8} {\tilde Q}_K {\tilde Q}_3 }& 0  \\
0 & {\displaystyle  { g^{'2}_1 \over 8} {\tilde Q}_{\bar K} {\tilde Q}_3 } \
\end{array} \right)  \ .
\end{eqnarray}

The $2 \times 2$ diagonalizing matrix for the scalar quarks, $U^{\tilde q}$, is given as
\begin{equation}
U^{\tilde q} =
\left(\begin{array}{cc}
\cos \theta_{\tilde q} & - \sin \theta_{\tilde q}  \\
\sin \theta_{\tilde q} & \cos \theta_{\tilde q}  \
\end{array} \right)  \ ,
\end{equation}
where the mixing angle $\theta_{\tilde q}$ is defined as
\begin{equation}
\tan \theta_{\tilde q} =  {m_{{\tilde q}_1}^2 - M_{11}^{\tilde q} \over M_{12}^{\tilde q}}  \  ,
\end{equation}
with $m_{{\tilde q}_1}$ being the mass of the lighter scalar quark
and $M_{ij}^{\tilde q}$ being the elements of the mass matrix $M^{\tilde q}$
for the scalar quarks (${\tilde q} = {\tilde t}, {\tilde b}, {\tilde k}$).

Now, we are ready to calculate the production cross section of the scalar Higgs bosons at the LHC.
In order to obtain $\sigma_i (pp \to S_i)$, the desired cross section at the laboratory level,
we should fold the cross section at the parton level with the gluon distribution functions as [25,26,22]
\begin{equation}
\sigma_i (E, m_{S_i})
= \int_{\tau_S}^1 d\tau \int^1_\tau dx {\tau_S \over x} g (x, m_{S_i}^2) g (\tau/x,m_{S_i}^2)
{\hat \sigma}_i ({\hat s} = \tau E^2 )   \   ,
\end{equation}
where $g(x, m^2_{S_i})$ is the gluon distribution function at the factorization scale $m^2_{S_i}$,
with $x$ being the fraction of the momentum of the proton carried by the gluon,
$g (\tau_S/x,m_{S_i}^2)$ is the gluon distribution functions for the other participating gluon,
and $\tau_S = m_{S_i}^2/E^2$ is the Drell-Yann variable,
with $E$ being the c.m. energy of the colliding protons at the LHC.
For the gluon distribution functions, the CTEQ6L is used in our calculation [28].

\section{Higgs Decays}

The scalar Higgs bosons would decay, among others, into pairs of fermions.
In the SM, the pairs of bottom quarks would be the most prominent decay modes for the SM Higgs boson
if its mass is below 135 GeV.
One may sort out the branching ratios for the SM Higgs boson with an intermediate mass of around 110 GeV as
$BR(H_{\rm SM} \to bb) > BR(H_{\rm SM} \to \tau \tau) > BR(H_{\rm SM} \to gg)$.
Thus, for such an intermediate SM Higgs boson, decaying into a pair of gluons via the triangular loop
of the top quark would be more difficult than decaying into a pair of bottom quarks or a pair of tau leptons.
On the other hand, if the mass of the SM Higgs boson turns out to be larger than 135 GeV,
the Higgs decay into a pair of gauge bosons would be more dominant than any other decay modes.

The situation in the USSM, we expect, would be different from the SM case,
since there are exotic quark and exotic scalar quarks in the USSM.
In particular, they would certainly affect the decays of the scalar Higgs bosons.
As one can see in the superpotential of the USSM, the heaviest scalar Higgs boson couples directly to the exotic quarks,
while the other scalar Higgs bosons couple indirectly to the exotic quarks through the mixings among them.
Thus, in order to study the effects of the exotic quarks in the decays of the scalar Higgs bosons,
we may as well pay more attentions to the heaviest scalar Higgs boson than the other ones.
Let us study the decay modes of the scalar Higgs bosons of the USSM into pairs of fermions,
gluons, and gauge bosons, as they are considered to be the most important decay modes.
Among them, the decay mode into gluon pairs should be studied in more detail,
since it may reveal the effects of the exotic quarks more clearly than other decay modes.

The total decay width of the $S_i$ ($i = 1,2,3$) may be assumed as
\begin{eqnarray}
\Gamma(S_i) & = & \Gamma (S_i \to b b) + \Gamma (S_i \to \tau \tau)
+ \Gamma (S_i \to \mu \mu) + \Gamma (S_i \to c c) \cr
& &\mbox{} + \Gamma (S_i \to s s) + \Gamma (S_i \to gg)
+ \Gamma (S_i \to WW) + \Gamma (S_i \to ZZ)  \ ,
\end{eqnarray}
where each decay mode can be understood without difficulty.
Note that the heaviest scalar Higgs boson, $S_3$, has additional decay modes,
since it can decay into the lighter ones, that is,
$S_3 \to S_1 S_1$, $S_3 \to S_2 S_2$, or $S_3 \to S_1 S_2$.

Now, let us describe each partial decays.
The partial decay width of $S_i$ into a pair of fermions is given as
\begin{equation}
\Gamma (S_i \to f f) = {C_f m_f^2 m_{S_i} \over 16 \pi v^2}
\left [ 1 - {4 m_f^2 \over m_{S_i}^2} \right ]^{3 \over 2} G_{S_i f f}^2 \ ,
\end{equation}
where $C_f$ is the color factor of fermions,
$m_f$ is the mass of the fermion, and $G_{S_i f f}$ is the coupling coefficient normalized with respect to
the corresponding SM coupling coefficient.
This fermionic partial decay width applies to bottom, charm, and strange quarks, and tau and muon leptons.

The partial decay width of $S_i$ into a pair of gauge bosons is given as
\begin{equation}
\Gamma (S_i \to VV) = {C_V m_{S_i}^3 \over 64 \pi v^2} \sqrt{1- 4 {m_V^2 \over m_{S_i}^2 } }
\left (1 - 4 {m_V^2 \over m_{S_i}^2 } + 12 {m_V^4 \over m_{S_i}^4 }  \right ) G_{S_iVV}^2  \ ,
\end{equation}
where $C_Z = 1$ and $C_W = 2$,
$m_V$ is the mass of the gauge boson, and $G_{S_i VV}$ is the coupling coefficient of $S_i$ to the gauge
boson pairs, normalized by the corresponding SM coupling coefficient ($V = Z, W$).

The normalized coupling coefficient of $S_i$ to a pair of $Z$ bosons is given as [21]
\begin{equation}
G_{S_i ZZ} = \cos \beta O_{1i} C_1^2 + \sin \beta O_{2i} C_2^2 + {s O_{3i} \over 4 G_{HZZ}} C_3^2 \ ,
\end{equation}
where $G_{HZZ} = g_2 m_Z/\cos \theta_W$ is the corresponding SM coupling coefficient,
with $\theta_W$ being the weak mixing angle,
and $C_i$ ($i = 1,2,3$) are dimensionless parameters defined as
\begin{eqnarray}
C_1 & = & \cos^2 \theta_W + \sin^2 \theta_W \cos \phi
- {g'_1 {\tilde Q}_1 \over g_2} \cos \theta_W \sin \theta_W \sin \phi \ , \cr
C_2 & = & \cos^2 \theta_W + \sin^2 \theta_W \cos \phi
+ {g'_1 {\tilde Q}_2 \over g_2} \cos \theta_W \sin \theta_W \sin \phi \ , \cr
C_3 & = & g'_1 {\tilde Q}_3 \sin \theta_W \sin \phi \ ,
\end{eqnarray}
with $\phi$ being the mixing angle between $Z$ and the extra neutral gauge boson in the USSM.
Notice that $\phi$ in $C_i$ induces the interference effect of the extra neutral gauge boson.

Also, the normalized coupling coefficient of $S_i$ to a pair of $W$ bosons is given as
\begin{equation}
G_{S_i WW} = \cos \beta O_{1i} + \sin \beta O_{2i}  \ .
\end{equation}
Note that it does not depend on $\phi$, since there is no extra charged gauge bosons in the USSM.

The partial decay width of $S_i$ into a pair of gluons is given as
\begin{equation}
\Gamma (S_i \to gg) = {\alpha_s^2 (m_Z) m_{S_i}^3 \over 64 \pi^3 v^2}  (| A^S  + A_D^S|^2)  \ ,
\end{equation}
where mediating loops involve top, bottom, and exotic quarks and their scalar partners.

Now, as we have mentioned before, the heaviest one of the three scalar Higgs bosons may
decay into the lighter members.
The partial decay width for $S_i \to S_j S_k$ is calculated as
\begin{equation}
\Gamma (S_i \to S_j S_k) = { C_S G_{S_j S_k S_i}^2  \over 16 \pi m_{S_i} }
\sqrt{\bigg (1 - {(m_{S_j} - m_{S_k})^2 \over m_{S_i}^2} \bigg )
\bigg (1 - {(m_{S_j} + m_{S_k})^2 \over m_{S_i}^2} \bigg )}  \ ,
\end{equation}
where $C_S$ is the symmetry factor and $G_{S_j S_k S_i}$ is the cubic Higgs coupling coefficient.
We have $C_S = 1$ for $j \neq k$ and $C_S = 1/2$ for $j = k$, and
\begin{eqnarray}
G_{S_j S_j S_i} & = & \lambda^2 s ( 2 O_{1j} O_{1i} O_{3j} + O_{1j}^2 O_{3i} + 2 O_{2j} O_{2i} O_{3j} + O_{2j}^2 O_{3i}) \cr
& &\mbox{} - \lambda A_{\lambda} (O_{1i} O_{2j} O_{3j} + O_{1j} O_{2i} O_{3j} + O_{1j} O_{2j} O_{3i}) \cr
& &\mbox{} + v \lambda^2 \cos \beta [O_{1i} (O_{2j}^2 + O_{3j}^2) + 2 O_{1j} (O_{2j} O_{2i} + O_{3j} O_{3i}) ] \cr
& &\mbox{} + v \lambda^2 \sin \beta [O_{2i} (O_{1j}^2 + O_{3j}^2) + 2 O_{2j} (O_{1j} O_{1i} + O_{3j} O_{3i}) ] \cr
& &\mbox{} + {v \over 4} (g_1^2 + g_2^2) \cos \beta [3 O_{1j}^2 O_{1i} - O_{1i} O_{2j}^2 - 2 O_{1j} O_{2j} O_{2i} )  ] \cr
& &\mbox{} + {v \over 4} (g_1^2 + g_2^2) \sin \beta [3 O_{2j}^2 O_{2i} - O_{2i} O_{1j}^2 - 2 O_{1j} O_{2j} O_{1i} )  ] \ , \cr
& &\mbox{} + g^{'2}_1 {\tilde Q}_3 s [2 O_{1j} O_{1i} O_{3j} {\tilde Q}_1 + O_{1j}^2 O_{3i} {\tilde Q}_1
+ 2 O_{2j} O_{2i} O_{3j} {\tilde Q}_2   \cr
& &\mbox{}  + O_{2j}^2 O_{3i} {\tilde Q}_2 + 3 O_{3j}^2 O_{3i} {\tilde Q}_3 ] \cr
& &\mbox{} + g^{'2}_1 v \cos \beta {\tilde Q}_1 [3 O_{1j}^2 O_{1i} {\tilde Q}_1 + O_{1i} O_{2j}^2 {\tilde Q}_2
+ 2 O_{1j} O_{2j} O_{2i} {\tilde Q}_2   \cr
& &\mbox{}  + O_{1i} O_{3j}^2 {\tilde Q}_3 + 2 O_{1j} O_{3j} O_{3i} {\tilde Q}_3 ] \cr
& &\mbox{} + g^{'2}_1 v \sin \beta {\tilde Q}_2 [2 O_{1j} O_{1i} O_{2j} {\tilde Q}_1 + O_{1j}^2 O_{2i} {\tilde Q}_1
+ 3 O_{2j}^2 O_{2i} {\tilde Q}_2   \cr
& &\mbox{}  + O_{2i} O_{3j}^2 {\tilde Q}_3 + 2 O_{2j} O_{3j} O_{3i} {\tilde Q}_3 ]    \ ,   \cr
G_{S_1 S_2 S_3} & = & \mbox{} - {1 \over 2} \lambda A_{\lambda}
(O_{13} O_{22} O_{31} + O_{12} O_{23} O_{31} + O_{13} O_{21} O_{32} \cr
& &\mbox{} + O_{11} O_{23} O_{32} + O_{12} O_{21} O_{33} + O_{11} O_{22} O_{33}) \cr
& &\mbox{} + \lambda^2 s ( O_{12} O_{13} O_{31} + O_{22} O_{23} O_{31} + O_{11} O_{13} O_{32} \cr
& &\mbox{} + O_{21} O_{23} O_{32} + O_{11} O_{12} O_{33} + O_{21} O_{22} O_{33} ) \cr
& &\mbox{} + v \lambda^2 \cos \beta [ O_{13} O_{21} O_{22} + O_{12} O_{21} O_{23} + O_{11} O_{22} O_{23} \cr
& &\mbox{} + O_{13} O_{31} O_{32} + O_{12} O_{31} O_{33} + O_{11} O_{32} O_{33} ] \cr
& &\mbox{} + v \lambda^2 \sin \beta [ O_{12} O_{13} O_{21} + O_{11} O_{13} O_{22} + O_{11} O_{12} O_{23} \cr
& &\mbox{} + O_{23} O_{31} O_{32} + O_{22} O_{31} O_{33} + O_{21} O_{32} O_{33} ] \cr
& &\mbox{} + {v \over 4}  (g_1^2 + g_2^2) \cos \beta [3 O_{11} O_{12} O_{13} - O_{13} O_{21} O_{22} \cr
& &\mbox{} - O_{12} O_{21} O_{23} - O_{11} O_{22} O_{23} ] \cr
& &\mbox{} + {v \over 4}  (g_1^2 + g_2^2) \sin \beta [3 O_{21} O_{22} O_{23} - O_{12} O_{13} O_{21} \cr
& &\mbox{} - O_{11} O_{13} O_{22} - O_{11} O_{12} O_{23} ] \cr
& &\mbox{} + g^{'2}_1 {\tilde Q}_3 s [O_{12} O_{13} O_{31} {\tilde Q}_1 + O_{11} O_{13} O_{32} {\tilde Q}_1
+ O_{11} O_{12} O_{33} {\tilde Q}_1   \cr
& &\mbox{} + O_{22} O_{23} O_{31} {\tilde Q}_2 + O_{21} O_{23} O_{32} {\tilde Q}_2
+ O_{21} O_{22} O_{33} {\tilde Q}_2 + 3 O_{31} O_{32} O_{33} {\tilde Q}_3 ] \cr
& &\mbox{} + g^{'2}_1 v \cos \beta {\tilde Q}_1 [3 O_{11} O_{12} O_{13} {\tilde Q}_1 + O_{13} O_{21} O_{22} {\tilde Q}_2
+ O_{12} O_{21} O_{23} {\tilde Q}_2   \cr
& &\mbox{} + O_{11} O_{22} O_{23} {\tilde Q}_2 + O_{13} O_{31} O_{32} {\tilde Q}_3 + O_{12} O_{31} O_{33} {\tilde Q}_3
+ O_{11} O_{32} O_{33} {\tilde Q}_3] \cr
& &\mbox{} + g^{'2}_1 v \sin \beta {\tilde Q}_2 [O_{12} O_{13} O_{21} {\tilde Q}_1 + O_{11} O_{13} O_{22} {\tilde Q}_1
+ O_{11} O_{12} O_{23} {\tilde Q}_1   \cr
& &\mbox{} + 3 O_{21} O_{22} O_{23} {\tilde Q}_2 + O_{23} O_{31} O_{32} {\tilde Q}_3 + O_{22} O_{31} O_{33} {\tilde Q}_3  \cr
& &\mbox{} + O_{21} O_{32} O_{33} {\tilde Q}_3]  \   .
\end{eqnarray}

\section{Numerical Analysis}

Now, we are ready for carrying out numerical analysis.
The scheme for setting the relevant parameters is similar to our previous analysis [18-22].
First of all, we note that there are strong experimental constraints in the USSM on the mass of
the extra gauge boson, $m_{Z'}$,
and on the size of the mixing angle $|\phi|$ between $Z$ and $Z'$.
In this article, we will use $m_{Z'} > 700$ GeV and $|\phi| < 3 \times 10^{-3}$.
The vacuum expectation value of the Higgs singlet is set as $s=700$ GeV in order to ensure a light $Z'$ scenario with a mass below 1 TeV.
The effective $U(1)'$ hypercharges of the Higgs doublets and Higgs singlet can be redefined as
$Q_i = g'_1 {\tilde Q}_i$ because ${\tilde Q}_i$ appear always together with ${g'}_1$.
Then, we fix the redefined effective hypercharges as $Q_1=-1$, $Q_2=-0.1$, and $Q_3 = 1.1$.
The Tevatron Run I data have set the lower bound on the mass of an long-lived quark with electric charges $\pm 1/3$ as 119 GeV [29].
The search for long-lived charged massive particles at Tevatron Run I
put a more stringent experimental lower bound of 180 GeV at the 95 \% confidence level [30].
Thus, we take the mass of the exotic quark as 400 GeV without contradicting the Tevatron constraint.
Also, the mass of the pseudoscalar Higgs boson at the one-loop level, $m_A$, will be used as an input parameter.

The present experimental lower bound on the mass of the SM Higgs boson set by LEP2 data is about 114.5 GeV [31].
Since the scalar Higgs bosons in the USSM are mixtures of the real components of the three Higgs fields,
the possibility of discovering them is affected by how much the real component of the Higgs singlet is mixed in them.
Further, the decay channel for the scalar Higgs bosons in the USSM is not restricted to a pair of bottom quarks.
Thus, the search for a USSM Higgs boson would depend on not only its mass but also its coupling coefficients and other factors.
Therefore, in our numerical analysis, the lower bound on the mass of the lightest scalar Higgs boson in the USSM
may not be restricted by the LEP2 constraint, although it does not imply that our numerical analysis may
phenomenologically contradict the experimental constraint of LEP2.
Our analysis have been performed by assuming that the scalar Higgs bosons
decay exclusively to a pair of bottom quarks.

It is well known that the top scalar quark plays a dominant role in the Higgs processes involving gluons, namely,
both the productions of a scalar Higgs boson via the gluon fusion process at the LHC and its decays into a gluon pair,
in the parameter region where the top scalar quark mass is below 400 GeV [25,26].
The result depends crucially on the mass of the top scalar quark.
In the USSM, as we concentrate on the role of the exotic quark and the exotic scalar quarks, the masses of
the exotic scalar quarks should be carefully studied.
Thus, it is of importance to investigate the dependence of the exotic scalar quark masses on the
trilinear soft SUSY breaking parameter, $A_k$, and on the mass of the exotic quark, $m_k$.

We show the result for the exotic scalar quark masses as functions of $A_k$ in Fig. 1, for $\tan \beta = 3$,
$\lambda = 0.387$, $m_Q = m_U = m_D = m_K = m_{\bar K} = 300$ GeV, $A_t = A_b = 440$ GeV, and $m_k = 400$ GeV and $m_A = 150$ GeV.
The mass of the lighter exotic scalar quark is calculated to be between 176 GeV and 428 GeV
while the mass of the heavier one is between 493 GeV and 628 GeV.
Their masses should be degenerate if $A_k$ vanishes, but actually they are not exactly degenerate in
mass since the non-zero $\lambda$ invokes a mixing between them.

Note that the masses of the other scalar quarks as well as the scalar Higgs masses do not depend on $A_k$.
The masses of the other scalar quarks are calculated as $m_{{\tilde t}_1} = 229$ GeV,
$m_{{\tilde t}_2} = 435$ GeV, $m_{{\tilde b}_1} = 252$ GeV, and $m_{{\tilde b}_2} = 339$ GeV.
The masses for the three scalar Higgs bosons at the one-loop level are calculated as
$m_{S_1} = 117$ GeV, $m_{S_2} = 156$ GeV, and $m_{S_3} = 783$ GeV, which are also shown in Fig. 1.

In order to study the effects of the exotic quark sector in the production of the scalar Higgs bosons in the
USSM via the gluon fusion process at the LHC,
it would be convenient to calculate the production cross sections with or without the contributions of the
exotic quark sector.
Let us denote the production cross section of $S_i$ as $\sigma^k (S_i)$ where only the exotic contributions are taken into account,
and the production cross section of $S_i$ as $\sigma^{t,b} (S_i)$ where they are not included.
These cross sections are obtained technically by taking either $A^S = 0$ or $A^S_D = 0$
in the formula for the production cross section.
We also introduce $\sigma (S_i)$ where the loops of top, bottom, and exotic quarks and their scalar partners are all included.
It should be noted that $\sigma (S_i)$ is not exactly the sum of $\sigma^{t,b} (S_i)$ and $\sigma^k (S_i)$,
since there are interference terms in $|A^S + A^S_D|^2$ in the formula for the production cross section.
However, the interference between them are negligible, for the parameter values we consider,
such that $\sigma (S_i)$ is almost equal to $\sigma^{t,b} (S_i) + \sigma^k (S_i)$.
The results are shown in Figs. 2a and 2b, where the values of the relevant parameters are the same as in Fig. 1.

In Fig. 2a, $\sigma^{t,b} (S_i)$ and $\sigma^k (S_i)$ are plotted as functions of $A_k$.
Note that $\sigma^{t,b} (S_i)$ do not change against $A_k$ because both the scalar Higgs boson masses
and the relevant couplings are independent of $A_k$, whereas the dependency on $A_k$ of $\sigma^k (S_i)$ is very clear.
Note also that $\sigma^{t,b} (S_3)$ is negligibly smaller than either $\sigma^{t,b} (S_2)$ or $\sigma^{t,b} (S_1)$.
This is mainly because the mass of $S_3$ is much heavier than the other two scalar Higgs bosons.
The trend is consistent with the SM prediction that the production cross section decreases as the Higgs mass increases.
It is also worthwhile noticing that $\sigma^k (S_3)$ is about $10^3$ times larger than $\sigma^{t,b} (S_3)$.
This fact indicates that the loops of exotic quark and exotic scalar quarks virtually dominate the $S_3$ productions
through the gluon fusion process in $pp$ collisions.

In Fig. 2b, $\sigma (S_i)$ ($i = 1,2,3$) are plotted as functions of $A_k$.
Comparing with Fig. 2a, one may note that $\sigma (S_1) \cong \sigma^{t,b} (S_1)$, $\sigma (S_2) \cong \sigma^{t,b} (S_2)$,
but $\sigma (S_3) \cong \sigma^k (S_3)$, in particular if $A_k$ is not so much large as 400 GeV.
Therefore, we may expect that the productions of the heaviest scalar Higgs bosons of the USSM at the LHC
would definitely exhibit the contributions of exotic quark and exotic scalar quarks in the USSM.

Now, let us study decay processes.
We calculate the partial decay width of $S_i$ ($i =1,2,3$) into a pair of gluons.
Here, too, it would be convenient to separate the contributions of exotic quark and exotic scalar quark loops from
the contributions of top and bottom quark and scalar quark loops.
Let us denote the partial decay width of $S_i$ into a gluon pair as $\Gamma^k (S_i)$ where only the exotic
contributions are taken into account,
and the partial decay width of $S_i$ as $\Gamma^{t,b} (S_i)$ where they are not included.
We also introduce $\Gamma (S_i)$ where the loops of top, bottom, and exotic quarks and their scalar partners are all included.

We show $\Gamma^{t,b} (S_i)$ and $\Gamma^k (S_i)$ in Fig. 3a. They are plotted as functions of $A_k$.
The other parameter values are the same as in Fig. 1.
Like $\sigma^{t,b} (S_i)$, one can see that $\Gamma^{t,b} (S_i)$ are all constant against $A_k$.
Note that $\Gamma^{t,b} (S_1) > \Gamma^{t,b} (S_2) >\Gamma^{t,b} (S_3)$, but the difference is not significant.
On the other hand, there is a great hierarchy among $\Gamma^k (S_i)$.
Note that $\Gamma^k (S_3)$ is larger than either $\Gamma^k (S_1)$ or $\Gamma^k (S_2)$ by several orders of magnitude.

In Fig. 3b, $\Gamma (S_i)$ ($i =1,2,3$) are plotted as functions of $A_k$.
One may notice that $\Gamma (S_i) \neq \Gamma^{t,b} (S_i) + \Gamma^k (S_i)$.
This is due to the interferences between them.
Clearly, by comparing Fig. 3a and Fig. 3b, the discrepancy between $\Gamma (S_1)$ and $\Gamma (S_2)$ is more visible than the discrepancy
between $\Gamma^{t,b} (S_1)$ and $\Gamma^{t,b} (S_2)$.
We expect that the interference between $\Gamma^{t,b} (S_2)$ and $\Gamma^k (S_2)$ is rather destructive such that
$\Gamma (S_2) < \Gamma^{t,b} (S_2)$.
Nevertheless, for the heaviest scalar Higgs boson, it is evident that practically $\Gamma (S_3) \cong \Gamma^k (S_3)$,
since $\Gamma^k (S_3) \gg \Gamma^{t,b} (S_3)$.

Next, we turn to the branching ratios of $S_i$ ($i = 1,2,3$).
The results are shown in Figs. 4-6.
The values of the relevant parameters are the same as in Fig. 1.
In Fig. 4, we show the branching ratios of $S_1$, namely, BR($S_1 \to bb$), BR($S_1 \to cc$), BR($S_1 \to \tau\tau$),
and BR($S_1 \to gg$), as functions of $A_k$.
The other branching ratios are very small to be included in the figure.
Note that the branching ratios for pairs of ordinary quarks are flat against $A_k$,
whereas BR($S_1 \to gg$) increases very slightly as $A_k$ increases,
since the mass of the lighter exotic scalar quark decreases down to about the top quark mass as $A_k$ increases.
Comparing Fig. 4 with the following figures, one can see that BR($S_1 \to bb$) is much larger than BR($S_2 \to bb$),
or BR($S_3 \to bb$) that is too small to be shown in Fig. 6.

In Fig. 5, we show the branching ratios of $S_2$, namely, BR($S_2 \to bb$), BR($S_2 \to cc$), BR($S_2 \to \tau\tau$),
and BR($S_2 \to gg$), as functions of $A_k$.
Here, one may notice that BR($S_2 \to cc$) $>$ BR($S_2 \to bb$), and BR($S_2 \to gg$) $>$ BR($S_2 \to \tau\tau$), while
the corresponding inequalities are reversed in case of $S_1$, as shown in Fig. 4.
Also, notice that BR($S_2 \to gg$) decreases mildly as $A_k$ increases by the contribution of the exotic quark and
exotic scalar quarks for the full range of $A_k$.

Since $S_3$ is much heavier than $S_1$ or $S_2$, its decay channels are richer than the lighter scalar Higgs bosons.
In particular, it may decay into a pair of exotic scalar quarks or into a pair of exotic quarks,
depending on the masses of participating particles.
According to our analysis, the masses are obtained as $m_{S_3}$= 780 GeV, $m_k$ = 400 GeV,
and $200 < m_{{\tilde k}_1} < 450$ GeV.
Thus, $S_3$ may decay into a pair of exotic quarks off shell,
and into a pair of the lighter exotic scalar quarks both on and off shell.

Let us consider the on-shell decays of $S_3$ into a pair of the lighter exotic scalar quarks for $A_k > 150$ GeV.
Its partial decay width is given by
\[
\Gamma (S_3 \to {\tilde k}_1 {\bar {\tilde k}}_1 ) = { 3 v^2 \over 8 \pi m_{S_3} }
(G_{S_3 {\tilde k}_1 {\tilde k}_1 })^2
\sqrt{ 1 - 4 {m_{{\tilde k}_1}^2 \over m_{S_3}^2 } }
\]
where $G_{S_3 {\tilde k}_1 {\tilde k}_1 }$ is given by Eq. (29).
This formula is used to calculate the branching ratio of $S_3$ into a pair of ${\tilde k}_1$.
In Fig. 6, we show BR($S_3 \to {\tilde k}_1 {\tilde k}_1 $),
the branching ratio for $S_3$ decays into a pair of the lighter exotic scalar quarks,
as function of $A_k$,
as well as other dominant branching ratios of $S_3$, namely,
BR($S_3 \to S_1S_1$), BR($S_3 \to S_2 S_2$), BR($S_3 \to S_1 S_2$), BR($S_3 \to tt$),
BR($S_3 \to WW$), BR($S_3 \to ZZ$), and BR($S_3 \to gg$).

It is clear in Fig. 6 that $S_3$ would decay mostly into a pair of lighter scalar Higgs bosons:
BR($S_3 \to S_1S_1$) and BR($S_3 \to S_2 S_2$) are almost equal and dominant over other decay channels.
Nevertheless, BR($S_3 \to {\tilde k}_1 {\bar {\tilde k}}_1 $) increases sharply at $A_k \sim 150$ GeV,
and becomes quite compatible to them.
Note that BR($S_3 \to {\tilde k}_1 {\bar {\tilde k}}_1 $) reaches its maximum at $A_k = 270$ GeV.
One may also notice that the branching ratio for a gluon pair is far less than 1 \%,
and fluctuates visibly due to the $A_k$ dependence of the exotic contributions.
It is comparable to BR($S_3 \to WW$) or BR($S_3 \to ZZ$).
However, the $S_3$ decays into a gluon pair is almost solely mediated
by the loops of the exotic quark and exotic scalar quarks.

In Fig. 6, one may notice that $S_3$ may decay quite dominantly into a pair of $S_1$, a pair of $S_2$,
a pair of the lighter exotic scalar quarks, a pair of top quarks, or into $S_1$ and $S_2$, as well as a pair of gluons.
The branching ratio of $S_3$ into a pair of gluons is very small as 0.001 at best.
We can obtain the number of raw Higgs events by using the integrated luminosity of the detectors such
as CMS [32] or ATLAS [33].
For the integrated luminosity of 700 fb$^{-1}$ which is expected to be reached in about ten years at the
LHC (CMS + ATLAS), about 140 raw Higgs events for $S_3$ would be produced at the LHC [34].
Thus, the decay mode of $S_3$ into a pair of gluons would not be relevant for any practical purpose,
since the background for such events is large at the LHC.
Consequently, other decay modes would be more worthwhile studying, or the super LHC (SLHC),
an upgrade of the LHC, is necessary to study the gluon decays of the heaviest quark in our model.
Note that the target of the integrated luminosity for SLHC is about 3000 fb$^{-1}$ during
the period of its operation [34].

The mass of the SM Higgs boson is determined by the vacuum expectation value of the
electroweak symmetry breaking, in which itself plays a crucial role.
Thus, it is natural to expect that the SM Higgs boson has a mass of the
scale of the electroweak symmetry breaking.
By requiring that there should be no Landau pole up to the cut off scale,
the renormalization group equation for the mass of the SM Higgs boson
predicts that it is smaller than 190 GeV, if the cut off scale is the Planck mass scale.

If the SM Higgs boson is discovered with a mass of the electroweak symmetry breaking scale
and if $S_3$ of the present model is about 780 GeV,
there would be no serious difficulty to distinguish between them.
On the other hand, if the SM Higgs boson is discovered with a mass comparable to the mass of $S_3$ of
the present model, their decay modes may be a useful tool to distinguish between them.
For the parameter value we consider, the mass of $S_3$ is about 780 GeV, and its decay modes are
shown in Fig. 6.
Let us assume that the mass of the SM Higgs boson is also about 780 GeV.
With this mass, the SM Higgs boson would decay 60 \% in $W$ boson pairs, 30 \% in $Z$ boson pairs,
and 7 \% in top quark pairs.
On the other hand, as Fig. 6 shows, $S_3$ would decay 45 \% into $S_1 S_1$ and 45 \% into
$S_1 S_2$ for $0 < A_k {\rm ~(GeV)~}< 130$,
or 30 \% each for $S_1 S_1$, $S_1 S_2$, and ${\tilde k}_1 {\bar {\tilde k}_1}$
for $A_k \sim 250$ GeV.
Further, as Figs. 4 and 5 show, $S_1$ and $S_2$ would respectively decay into a pair of bottom quarks
by 60 \% and 19 \%.
Therefore, using the decay modes of bottom quark pairs as background, assuming 700 fb$^{-1}$ for
the integrated luminosity of the LHC, we would have
29862 events of $S_3 \rightarrow b{\bar b}$ whereas 2362 events for the same decay modes
for the SM Higgs bosons.
In this case, the signal significance is more than $614 \sigma$, well above the discovery limit.

Since $S_3$ is much heavier than $S_1$ or $S_2$, it may decay into a pair of exotic scalar quarks or into a pair of exotic quarks.
For the parameter values we take up to now, the mass of $S_3$ is 780 GeV, the mass of the exotic quarks is
400 GeV, and the mass of the lighter exotic scalar quarks is between 180 and 450 GeV.
Thus, for this choice of parameter values, $S_3$ may decay into a pair of the lighter exotic scalar quarks both on and off shell.
On the other hand, it is kinematically forbidden for $S_3$ to decay into a pair of exotic quarks on shell.

In order to study the on-shell decays of $S_3$ into a pair of exotic quarks, we need to explore other regions of the parameter space.
We would like to study the parameter region where $m_k =$ 400 GeV and $m_{S_3}$ is larger than 800 GeV.
We find that a representative set of the parameter values in this region is $\tan\beta = 3$, $\lambda = 0.5$,
$m_Q = m_U = m_D = m_K = m_{\bar K} = 500$ GeV, $A_t = A_b = 440$ GeV, $m_k = 400$ GeV,
$m_A = 900$ GeV, and $0 <A_k < 900$ GeV.
These parameter values yield $176 < m_{{\tilde k}_1} {\rm ~(GeV)~} <585$,
$635 < m_{{\tilde k}_2} {\rm ~(GeV)~} <845$,
$m_{S_1} \sim 114.5$ GeV, $m_{S_2} \sim 750.6$ GeV, and $m_{S_3} \sim 928.7$ GeV,
satisfying our requirement.
Fig. 7 shows the masses of these particles as functions of $A_k$.
Hence, with this choice of parameter values, $S_3$ may decay into a pair of exotic quarks on shell.

We calculate the production cross sections for the scalar Higgs bosons via gluon fusion process at the LHC.
The radiative corrections include the contributions from loops of top quarks, bottom quarks,
exotic quarks, and their scalar superpartners. The results are shown in Fig. 8.
The parameter values are the same as Fig. 7.
We then calculate the branching ratios of the scalar Higgs bosons.
The results for the branching ratios are shown in Figs. 9, 10, and 11, for the same parameter values as Fig. 7.

These figures may be compared with Figs. 4, 5, and 6, respectively.
Consider first Fig. 9, which shows the branching ratios of $S_1$.
The behavior of $S_1$ in Fig. 9 is not so much different from its behavior in Fig. 4, mainly because
its mass is roughly the same in both figures:
$m_{S_1} \sim 114.5$ GeV in Fig. 9 and $m_{S_1} \sim 117$ GeV in Fig. 4.

By contrast, the behavior of the branching ratios of $S_2$ in Fig.10 is strikingly different from that in Fig. 5,
since $m_{S_2} \sim 750$ GeV in Fig. 10 and $m_{S_2} \sim 156$ GeV in Fig. 5.
In particular, the branching ratio for a pair of lighter exotic scalar quarks,
BR ($S_2 \rightarrow {\tilde k}_1 {\bar {\tilde k}}_1$), can be seen in Fig. 10,
whereas it is absent in Fig. 5, as it is forbidden kinematically there.
The coupling coefficient of $S_2$ to a pair of ${\tilde k}_1 {\bar {\tilde k}}_1$ depends very severely
on $A_k$.
Thus, the branching ratio for a pair of lighter exotic scalar quarks varies wildly as $A_k$ changes.
Note that $S_2$ cannot decay into a pair of exotic quarks on shell, since its mass is smaller than
$2 m_k$, in both figures.

The heaviest scalar Higgs boson, $S_3$, on the other hand, may decay into a pair of exotic quarks on shell,
as shown in Fig. 11.
One may notice that the branching ratio of $S_3 \rightarrow k{\bar k}$ is considerably larger
than the branching ratios for gauge boson pair or gluon pair.
Therefore, for some parameter regions where $S_3$ is allowed kinematically to decay into a pair of exotic quarks,
the decays of the heaviest scalar quark into exotic quarks may become important to examine our model.
Also, the branching ratio for a pair of exotic scalar quarks in Fig. 11 varies more wildly than in Fig. 6.

\section{Conclusions}

In this article, we study the Higgs sector of the USSM at the one-loop level.
The radiative corrections from the loops of top, bottom and exotic quarks and their superpartners
are calculated for the scalar and pseudoscalar Higgs boson masses,
where we include the $D$-term contribution.
For the numerical analysis, we take $\tan \beta = 3$, $\lambda = 0.387$,
$m_Q = m_U = m_D = m_K = m_{\bar K} = 300$ GeV, $A_t = A_b = 440$ GeV,
and the mass of the exotic quark as $m_k = 400$ GeV and
the mass of the pseudoscalar Higgs boson as $m_A = 150$ GeV.
We assume that CP symmetry is conserved in the Higgs sector of the USSM.
We find that the exotic scalar quarks become nearly degenerate in mass for small $A_k$.

For the USSM Higgs phenomenology at the LHC, we study the production processes and decay modes
of the scalar Higgs bosons that involve gluons.
As the scalar Higgs bosons may be produced in $pp$ collisions via the gluon fusion process,
we calculate the production cross sections for the gluon fusion process, considering the loop contributions from
top, bottom, and exotic quarks and their superpartners.
We find that, for the heaviest scalar Higgs boson of the USSM, the gluon fusion process is mediated
virtually only by the loops of exotic quark and exotic scalar quarks.
So is the partial decay width of the heaviest scalar Higgs boson into a pair of gluons.
In other words, practically, the heaviest scalar Higgs boson couples to a pair of gluons only
through the loops of exotic quark and exotic scalar quarks.
Thus, the contributions from the loops of top and bottom quarks and their scalar partners are negligible in the
gluon fusion process of the heaviest scalar Higgs boson production as well as in its partial decay width into a gluon pair.
However, the partial decay width of the heaviest scalar Higgs boson into a pair of gluons is less than 1 \%, just
comparable to its partial decay width into a pair of gauge bosons.

We should note that our results are obtained for a particular set of parameter values,
allowing $A_k$ to vary between 0 and 450 GeV.
Thus, it may be hard to say that this set of parameter values represent the whole parameter space of the USSM.
We may have to search a much wider region of the parameter space of the USSM in order to obtain more predictive
results for the contributions of the exotic quark and exotic scalar quarks in the Higgs phenomenology.
However, qualitatively, our results suggest that the exotic quark and exotic scalar quarks play observable roles in
the productions and decays of the scalar Higgs bosons of the USSM at the LHC.
In order to examine the USSM at the LHC, in particular to examine the effects of the exotic quark sector in the Higgs
physics, the heaviest scalar Higgs boson may be one of the best windows,
as our calculations clearly suggests.

\section*{Acknowledgments}

S. W. Ham thanks Jae Sik Lee for valuable comments.
He thanks P. Ko for the hospitality at KIAS where a part of this work has been performed.
He would like to acknowledge the support from KISTI under
"The Strategic Supercomputing Support Program (No. KSC-2008-S01-0011)"
with Dr. Kihyeon Cho as the technical supporter.
This work is supported by the Korea Research Foundation Grant funded by the Korean Government
(MOEHRD, Basic Research Promotion Fund) (KRF-2007-341-C00010).



\vfil\eject


\setcounter{figure}{0}
\def\figurename{}{}%
\renewcommand\thefigure{FIG. 1}
\begin{figure}[t]
\begin{center}
\includegraphics[scale=0.6]{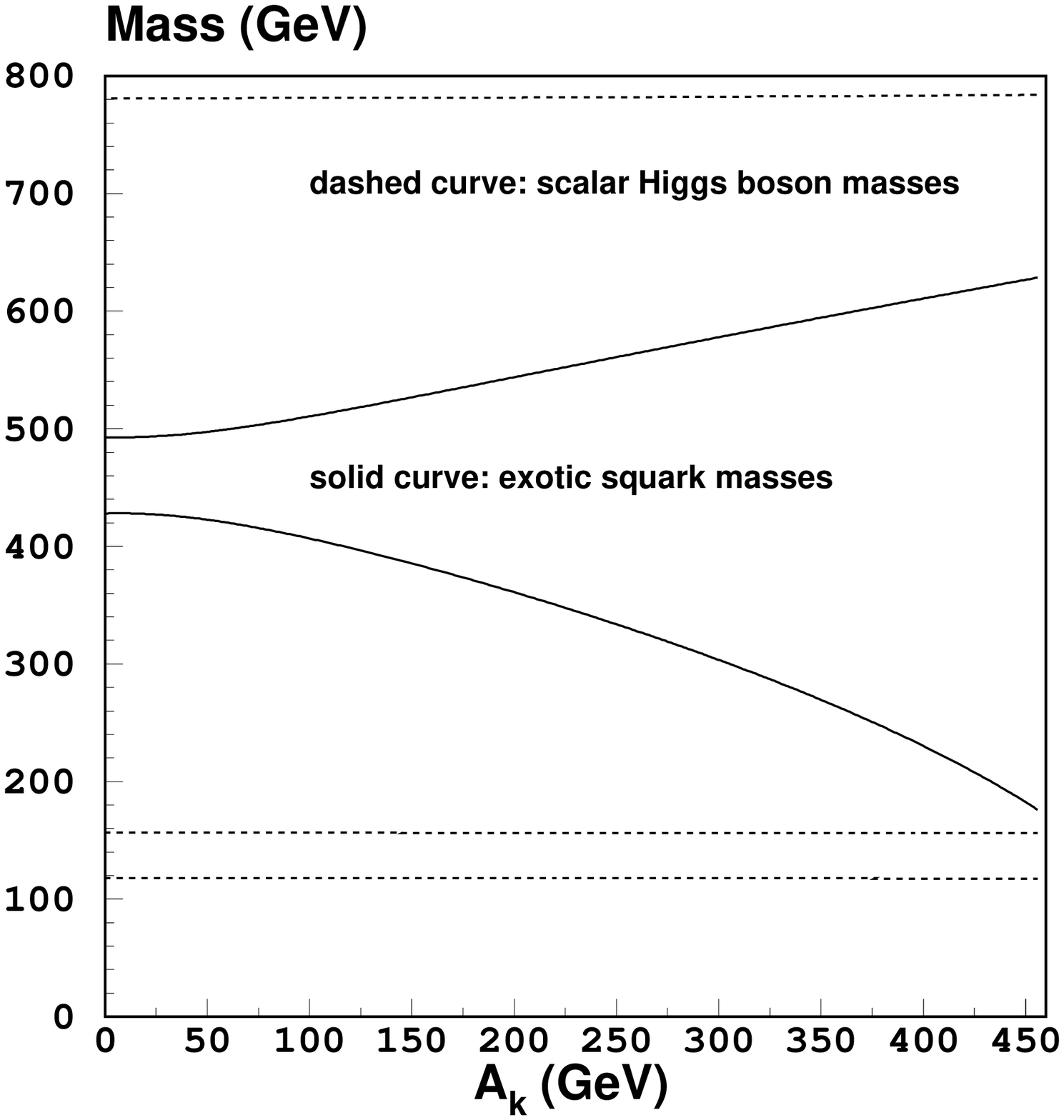}
\caption[plot]{The scalar Higgs boson masses $m_{S_i}$ ($i =1,2,3$) and the exotic scalar quark masses
$m_{{\tilde k}_j}$ ($j = 1,2$) are plotted as functions of $A_k$,
for $\tan \beta = 3$, $\lambda = 0.387$, $m_Q = m_U = m_D = m_K = m_{\bar K} = 300$ GeV,
$A_t = A_b = 440$ GeV, the mass of the pseudoscalar Higgs boson as $m_A = 150$ GeV,
and the mass of the exotic quark as $m_k = 400$ GeV.
Note that $m_{S_3} > m_{S_2} > m_{S_1}$ are independent of $A_k$,
whereas the gap between the masses of the two exotic scalar quarks widens as $A_k$ increases.}
\end{center}
\end{figure}

\setcounter{figure}{0}
\def\figurename{}{}%
\renewcommand\thefigure{FIG. 2a}
\begin{figure}[t]
\begin{center}
\includegraphics[scale=0.6]{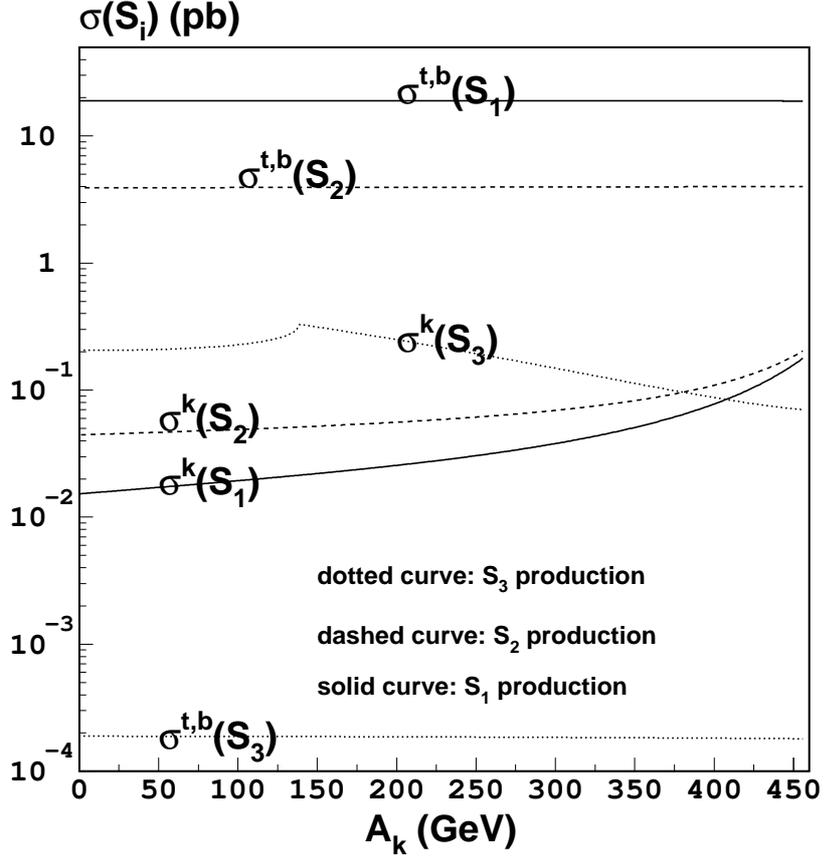}
\caption[plot]{The production cross sections of the scalar Higgs
bosons via the gluon fusion process at the LHC are plotted as functions of $A_k$.
$\sigma^{t,b} (S_i)$ ($i = 1,2,3$) are the cross sections where only the loops of
top quark, bottom quark, top scalar quarks, and bottom scalar quarks are taken into account,
and $\sigma^k (S_i)$ ($i = 1,2,3$) are the cross sections where only the loops of
the exotic quark and exotic scalar quarks are taken into account.
The parameter values are the same as Fig. 1.
Note that $\sigma^{t,b} (S_i)$ ($i = 1,2,3$) are independent of $A_k$. }
\end{center}
\end{figure}

\setcounter{figure}{0}
\def\figurename{}{}%
\renewcommand\thefigure{FIG. 2b}
\begin{figure}[t]
\begin{center}
\includegraphics[scale=0.6]{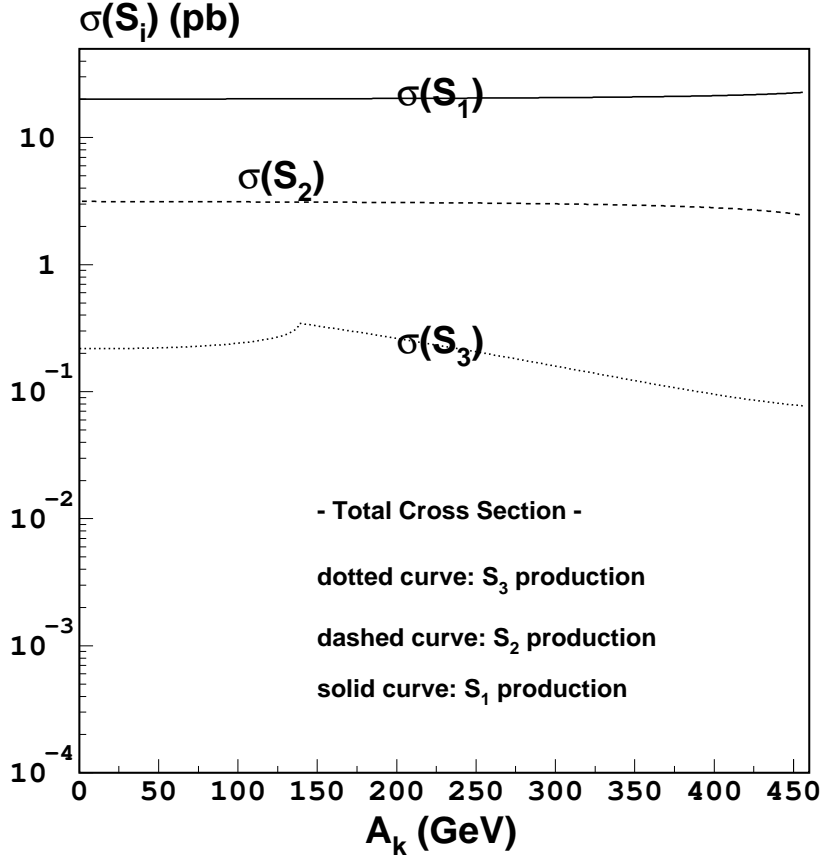}
\caption[plot]{The production cross sections of the scalar Higgs
bosons via the gluon fusion process at the LHC, $\sigma (S_i)$ ($i = 1,2,3$),
are plotted as functions of $A_k$.
They are calculated by taking into account together both the loops of
top quark, bottom quark, top scalar quarks, bottom scalar quarks and the loops of
the exotic quark and exotic scalar quarks.
The parameter values are the same as Fig. 1.
Notice that $\sigma (S_3)$ is virtually equal to $\sigma^k (S_3)$ in Fig. 2a.}
\end{center}
\end{figure}

\setcounter{figure}{0}
\def\figurename{}{}%
\renewcommand\thefigure{FIG. 3a}
\begin{figure}[t]
\begin{center}
\includegraphics[scale=0.6]{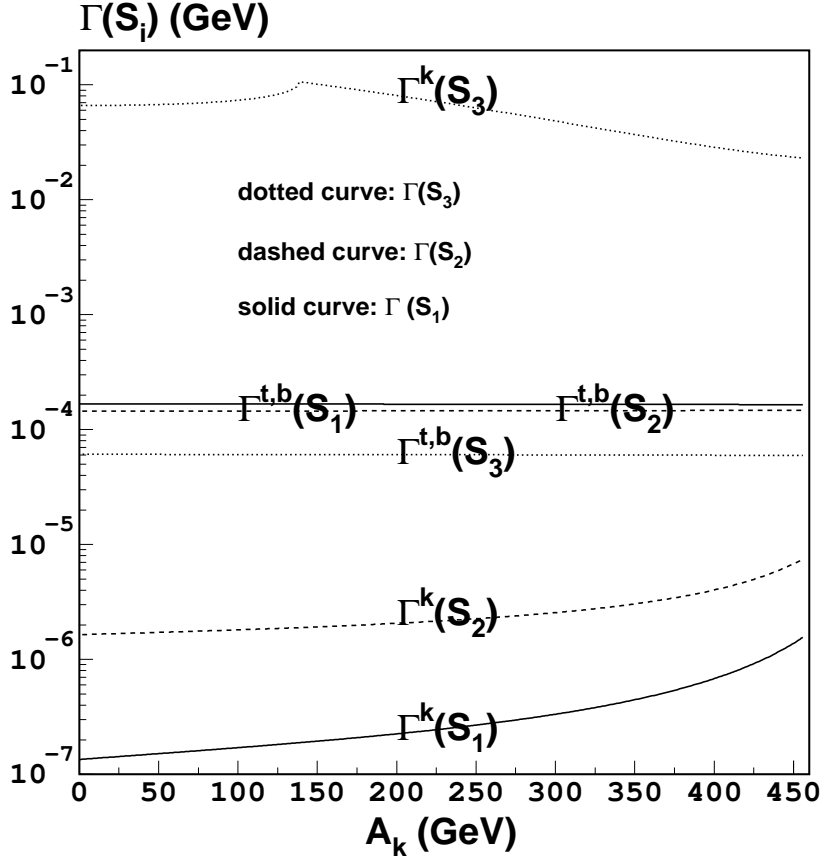}
\caption[plot]{The partial widths of the scalar Higgs boson decays
into a pair of gluons are plotted as functions of $A_k$.
$\Gamma^{t,b} (S_i)$ ($i = 1,2,3$) are the partial decay widths
where only the loops of top quark, bottom quark, top scalar quarks, and
bottom scalar quarks are taken into account,
and $\Gamma^k (S_i)$ are the decay widths where only the loops of
the exotic quark and exotic scalar quarks are taken into account.
The parameter values are the same as Fig. 1.
Note that $\Gamma^{t,b} (S_i)$ ($i = 1,2,3$) are independent of $A_k$.
Note also that $\Gamma^k (S_3)$ is much larger than any of $\Gamma^{t,b} (S_i)$
($i = 1,2,3$) whereas $\Gamma^k (S_1)$ or $\Gamma^k (S_2)$ are smaller than them.}
\end{center}
\end{figure}

\setcounter{figure}{0}
\def\figurename{}{}%
\renewcommand\thefigure{FIG. 3b}
\begin{figure}[t]
\begin{center}
\includegraphics[scale=0.6]{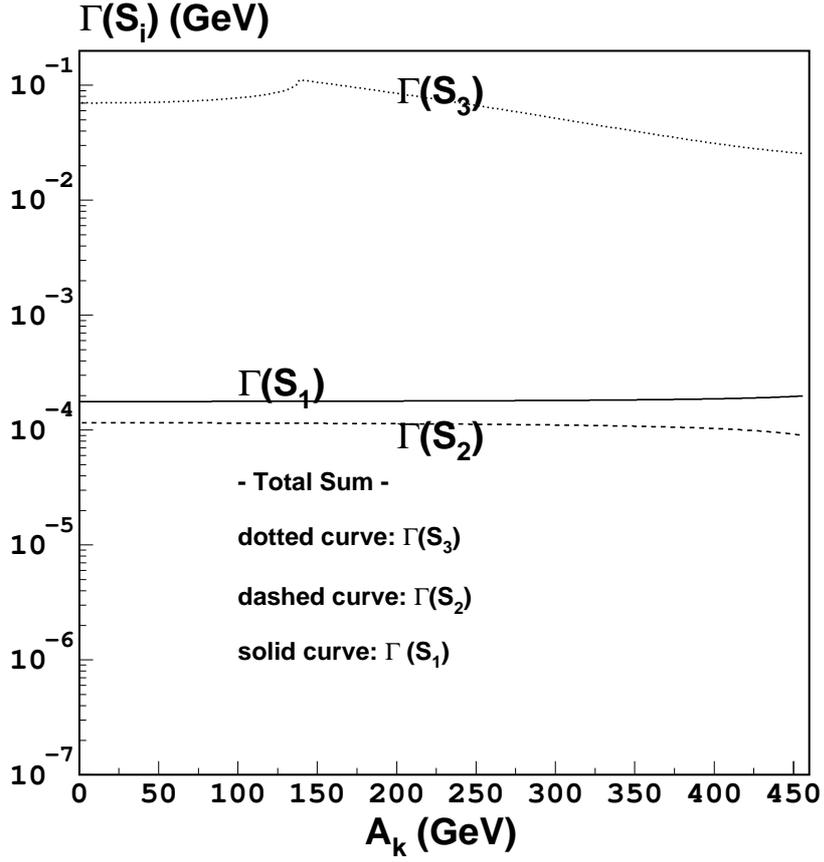}
\caption[plot]{The partial widths of the scalar Higgs boson decays
into a pair of gluons, $\Gamma (S_i)$ ($i = 1,2,3$) are plotted as functions of $A_k$.
They are calculated by taking into account together both
the loops of top quark, bottom quark, top scalar quarks, bottom scalar quarks,
and the loops of the exotic quark and exotic scalar quarks are taken into account.
The parameter values are the same as Fig. 1.
Note that $\Gamma (S_3)$ is practically identical to $\Gamma^k (S_3)$. }
\end{center}
\end{figure}

\setcounter{figure}{0}
\def\figurename{}{}%
\renewcommand\thefigure{FIG. 4}
\begin{figure}[t]
\begin{center}
\includegraphics[scale=0.6]{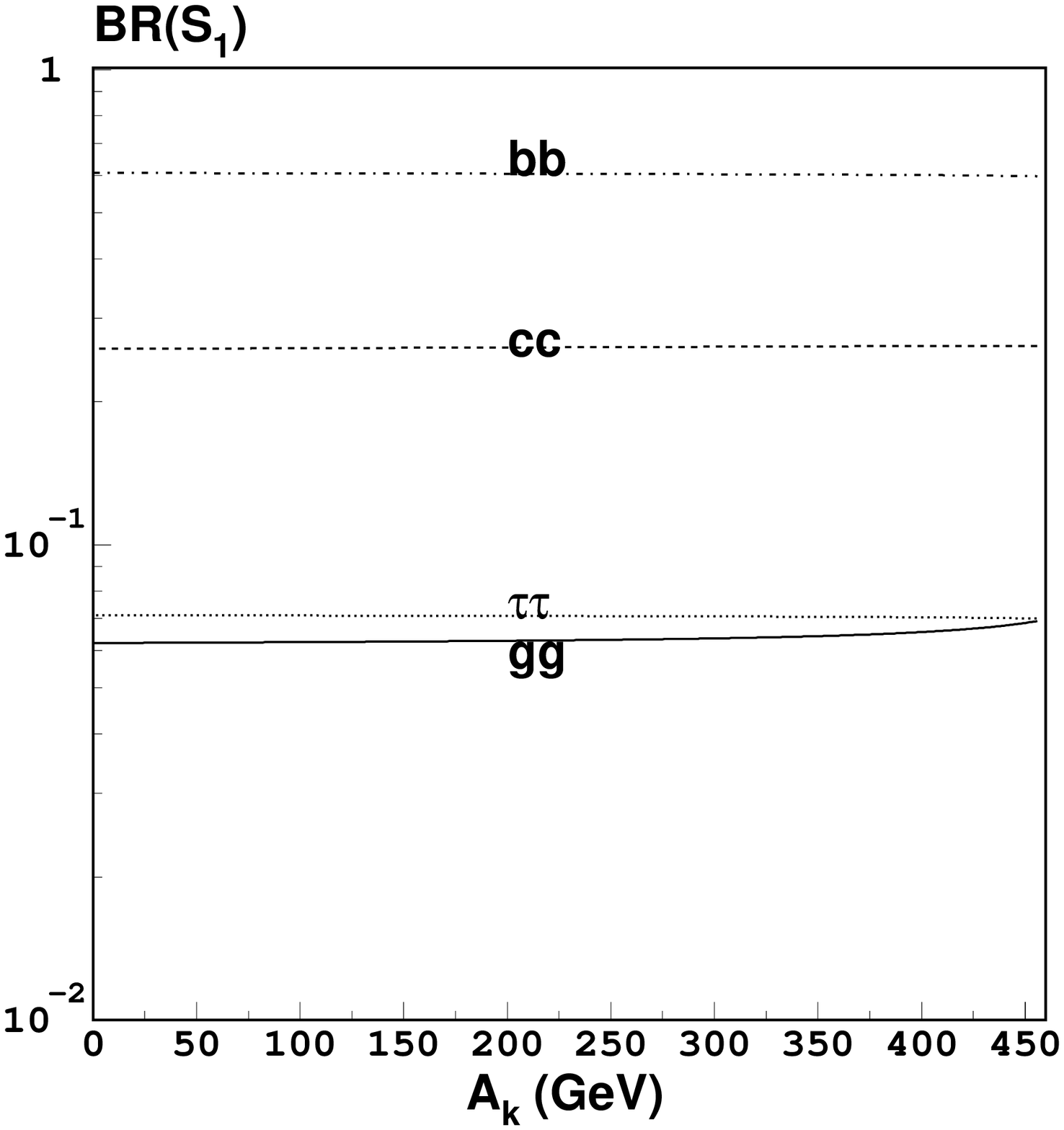}
\caption[plot]{The branching ratios of the lightest scalar Higgs boson, $S_1$, are plotted
as functions of $A_k$.
The parameter values are the same as Fig. 1.
Note that BR($S_1 \to bb$), BR($S_1 \to cc$), and BR($S_1 \to \tau\tau$) are
independent of $A_k$, whereas BR($S_1 \to gg$) slightly increases as $A_k$ increases.}
\end{center}
\end{figure}

\setcounter{figure}{0}
\def\figurename{}{}%
\renewcommand\thefigure{FIG. 5}
\begin{figure}[t]
\begin{center}
\includegraphics[scale=0.6]{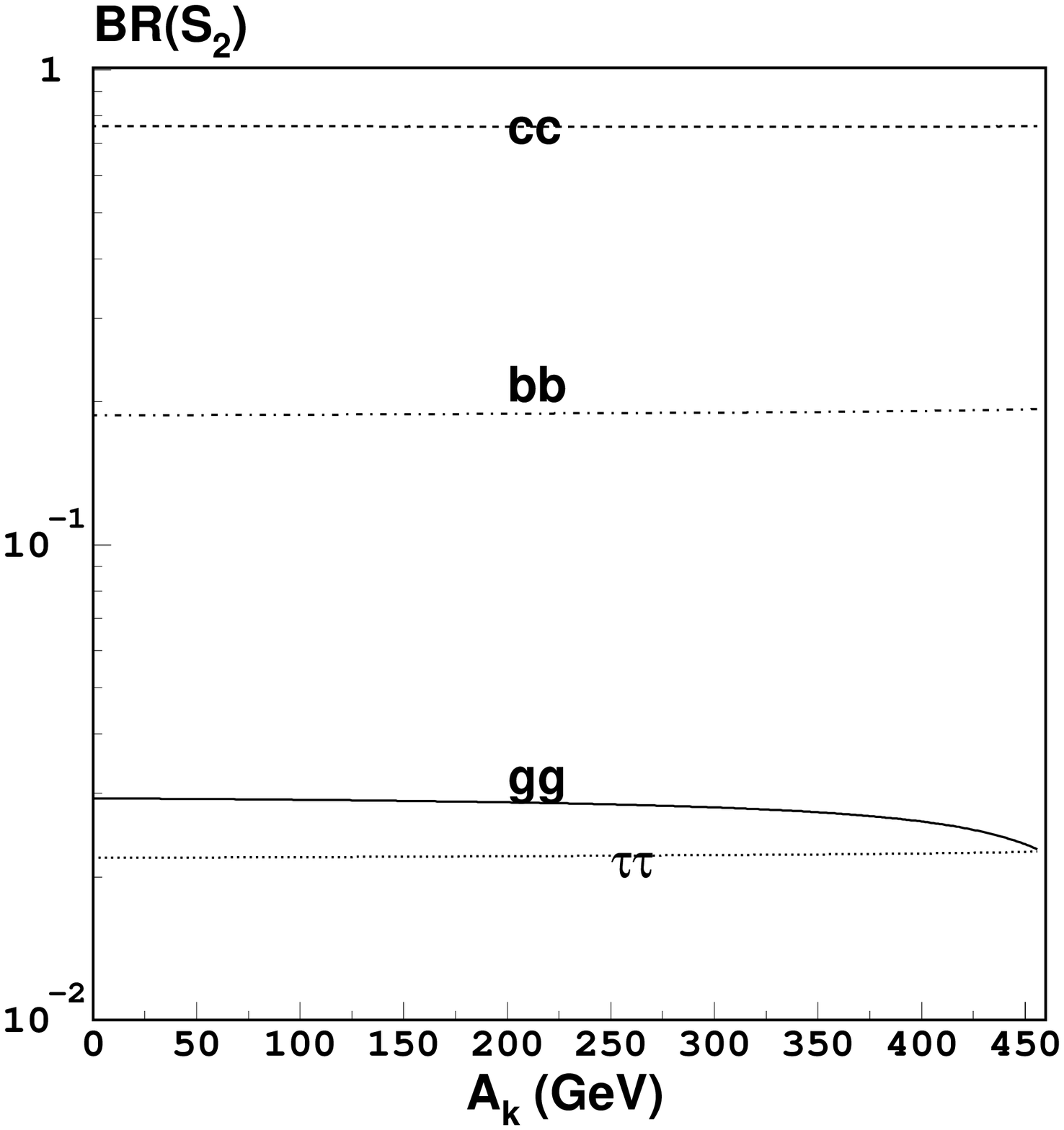}
\caption[plot]{The branching ratios of the middle heavy scalar Higgs boson, $S_2$, are plotted
as functions of $A_k$.
The parameter values are the same as Fig. 1.
Note that BR($S_2 \to bb$), BR($S_2 \to cc$), and BR($S_2 \to \tau\tau$) are
independent of $A_k$, whereas BR($S_2 \to gg$) slightly decreases as $A_k$ increases.
Note also that BR($S_2 \to gg$) is larger than BR($S_2 \to \tau\tau$). }
\end{center}
\end{figure}

\setcounter{figure}{0}
\def\figurename{}{}%
\renewcommand\thefigure{FIG. 6}
\begin{figure}[t]
\begin{center}
\includegraphics[scale=0.6]{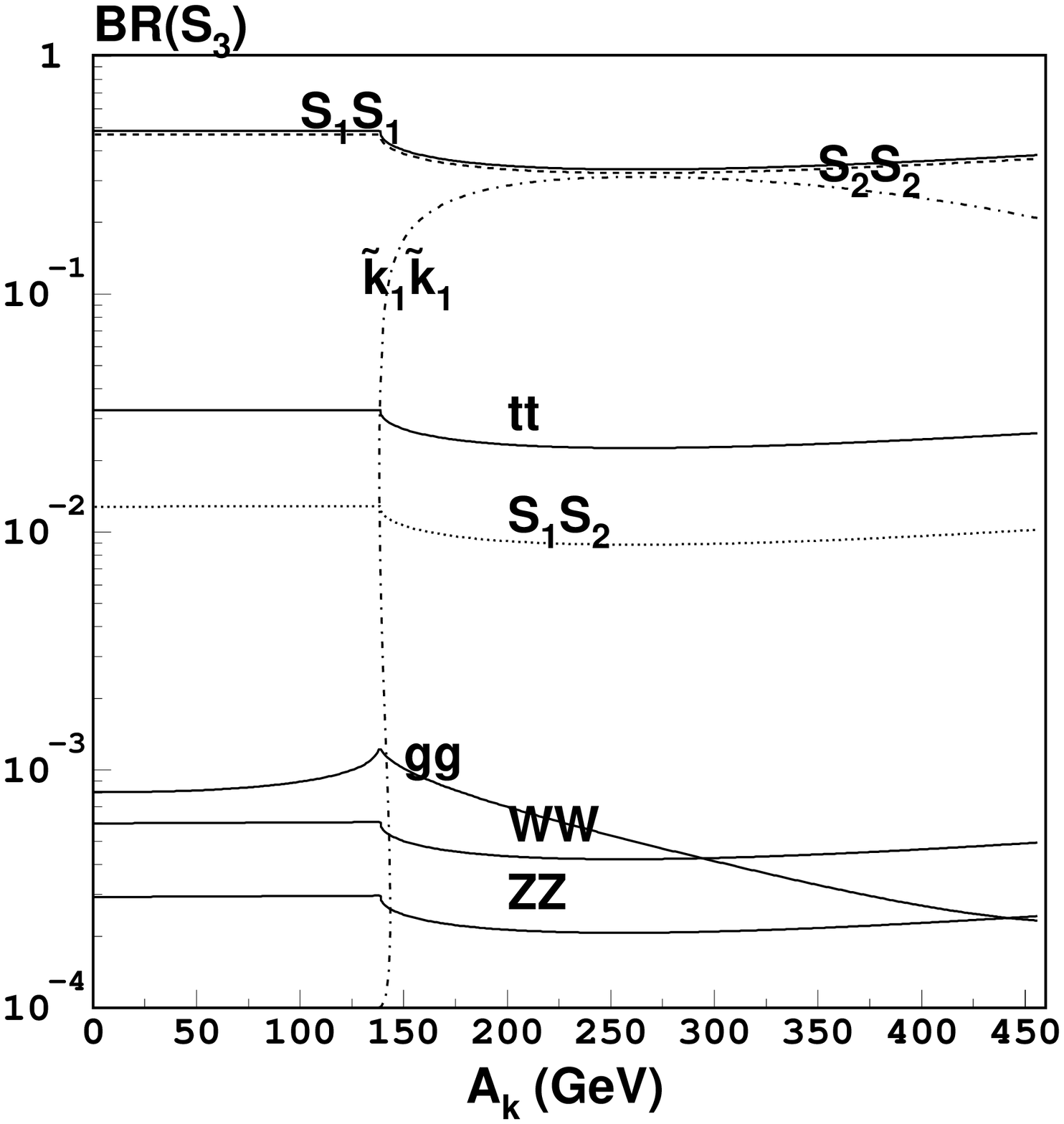}
\caption[plot]{The branching ratios of the heaviest scalar Higgs boson, $S_3$, are plotted
as functions of $A_k$.
The parameter values are the same as Fig. 1.
Note that BR($S_3 \to tt$), BR($S_3 \to WW$), BR($S_3 \to ZZ$), as well as
BR($S_3 \to S_1S_2$), BR($S_3 \to S_1S_1$), and BR($S_3 \to S_2S_2$), are
independent of $A_k$,
whereas BR($S_3 \to {\tilde k}_1 {\tilde k}_1$) and BR($S_3 \to gg$) depend significantly on $A_k$.
Note also that BR($S_3 \to gg$) is larger than or quite compatible to
BR($S_3 \to WW$) and BR($S_3 \to ZZ$). }
\end{center}
\end{figure}

\setcounter{figure}{0}
\def\figurename{}{}%
\renewcommand\thefigure{FIG. 7}
\begin{figure}[t]
\begin{center}
\includegraphics[scale=0.6]{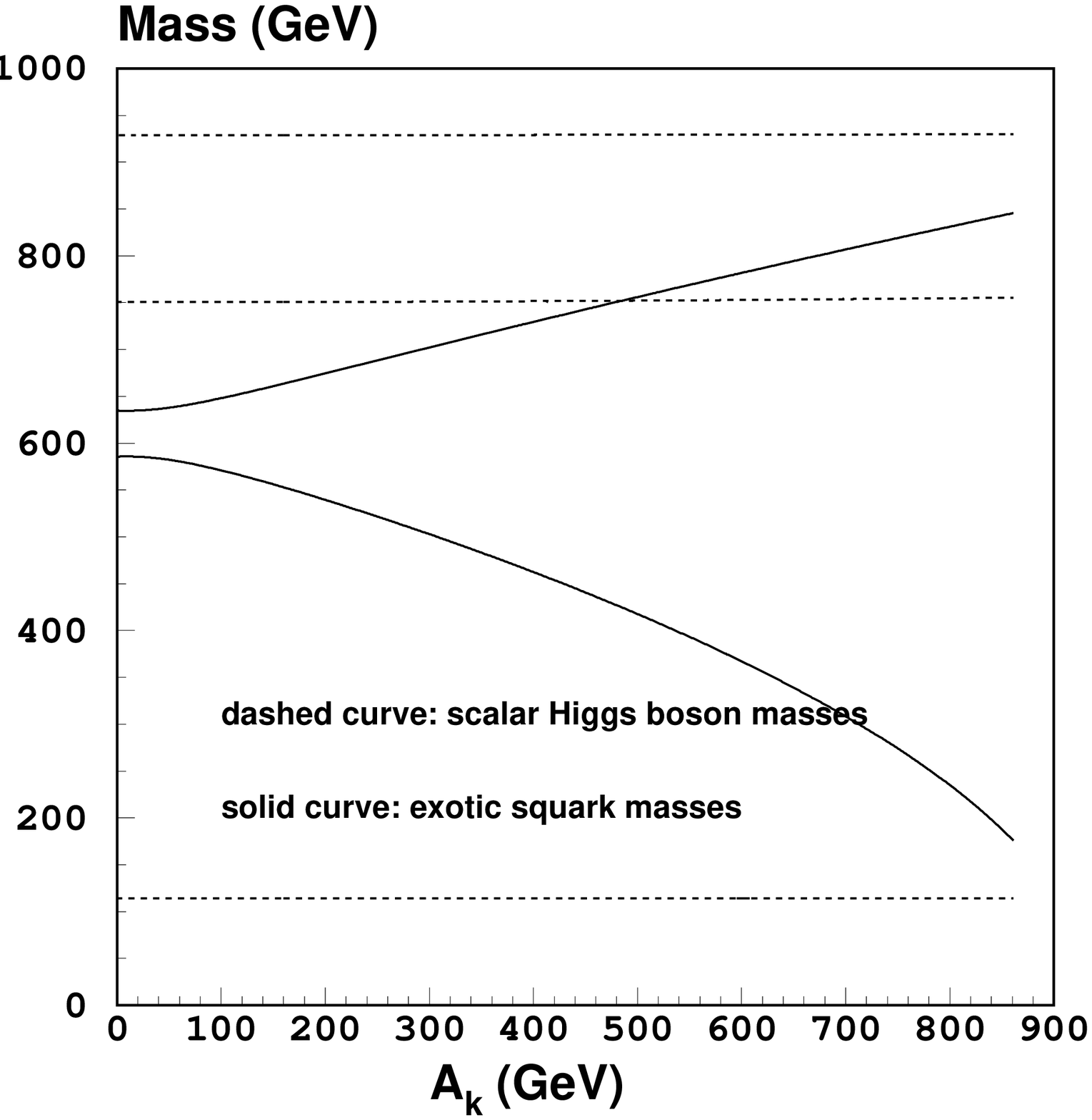}
\caption[plot]{The same as Fig. 1, except for a different set of parameter values:
 $\tan\beta = 3$, $\lambda = 0.5$,
$m_Q = m_U = m_D = m_K = m_{\bar K} = 500$ GeV, $A_t = A_b = 440$ GeV,
the mass of the pseudoscalar Higgs boson as $m_A = 900$ GeV,
and the mass of the exotic quark as $m_k = 400$ GeV.
Note that $m_{S_3}$ is larger than twice the mass of the exotic quark.}
\end{center}
\end{figure}

\setcounter{figure}{0}
\def\figurename{}{}%
\renewcommand\thefigure{FIG. 8}
\begin{figure}[t]
\begin{center}
\includegraphics[scale=0.6]{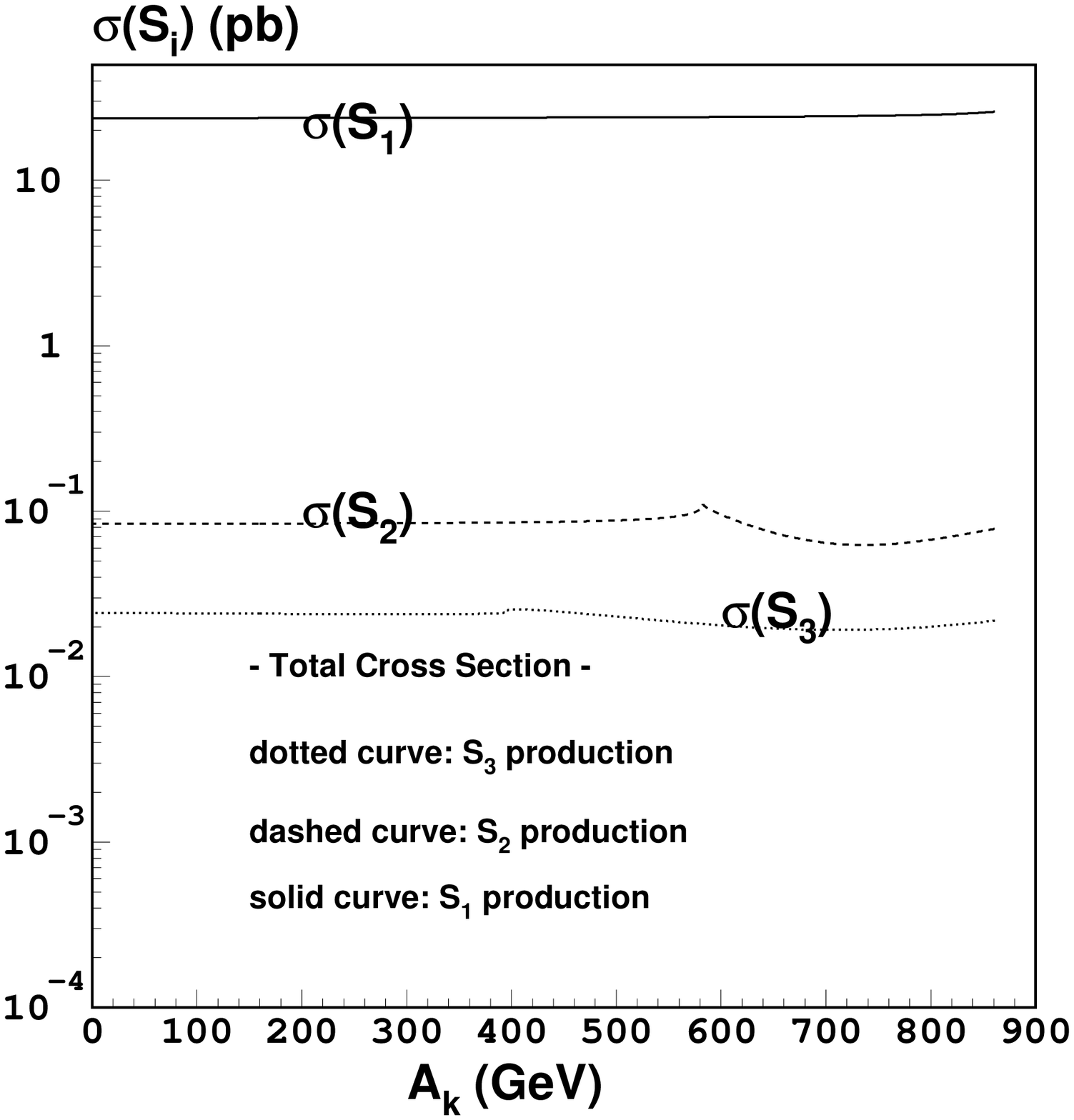}
\caption[plot]{The same as Fig. 2b, except that the parameter values are those of Fig. 7}
\end{center}
\end{figure}

\setcounter{figure}{0}
\def\figurename{}{}%
\renewcommand\thefigure{FIG. 9}
\begin{figure}[t]
\begin{center}
\includegraphics[scale=0.6]{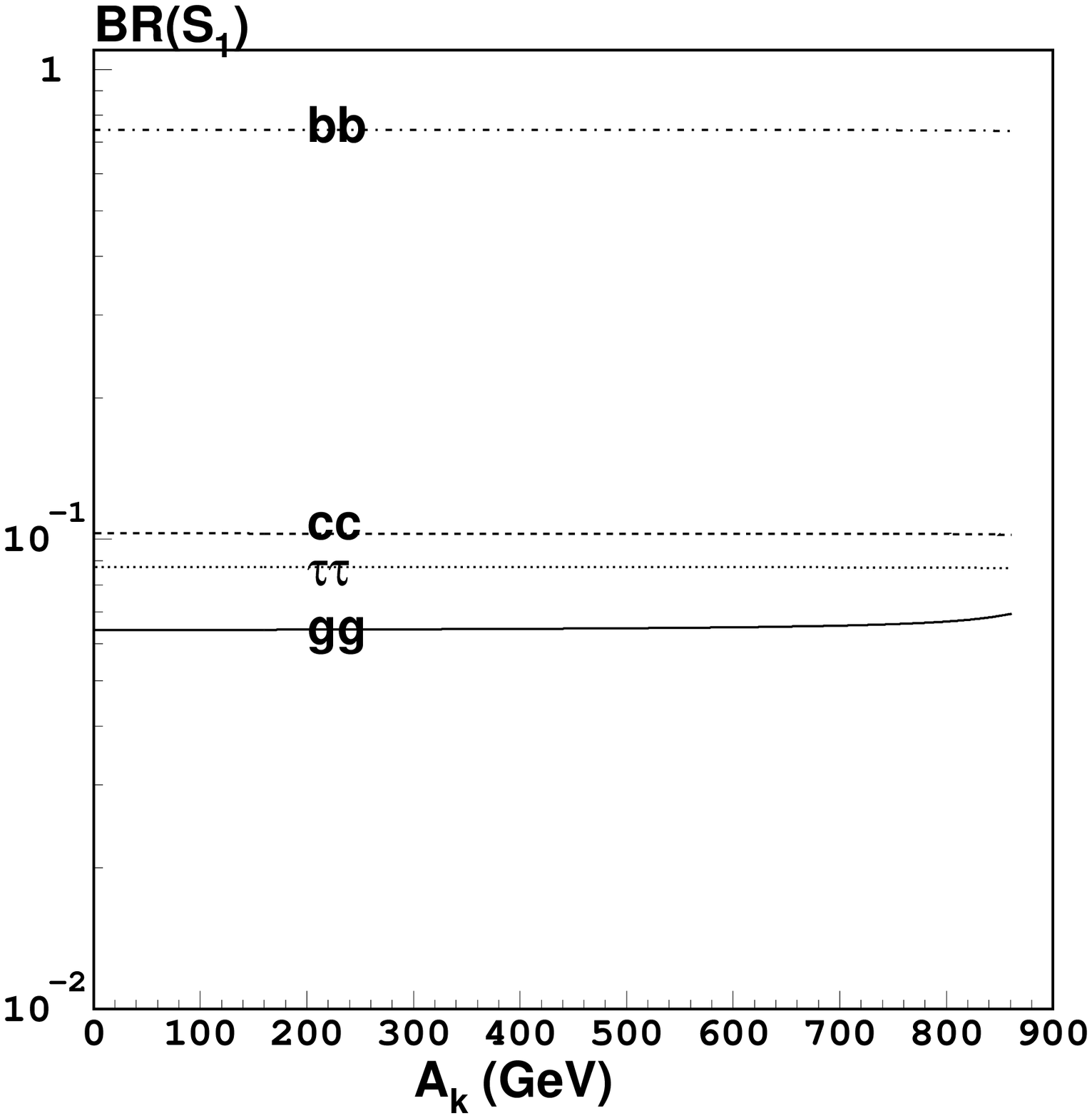}
\caption[plot]{The same as Fig. 4, except that the parameter values are those of Fig. 7}
\end{center}
\end{figure}

\setcounter{figure}{0}
\def\figurename{}{}%
\renewcommand\thefigure{FIG. 10}
\begin{figure}[t]
\begin{center}
\includegraphics[scale=0.6]{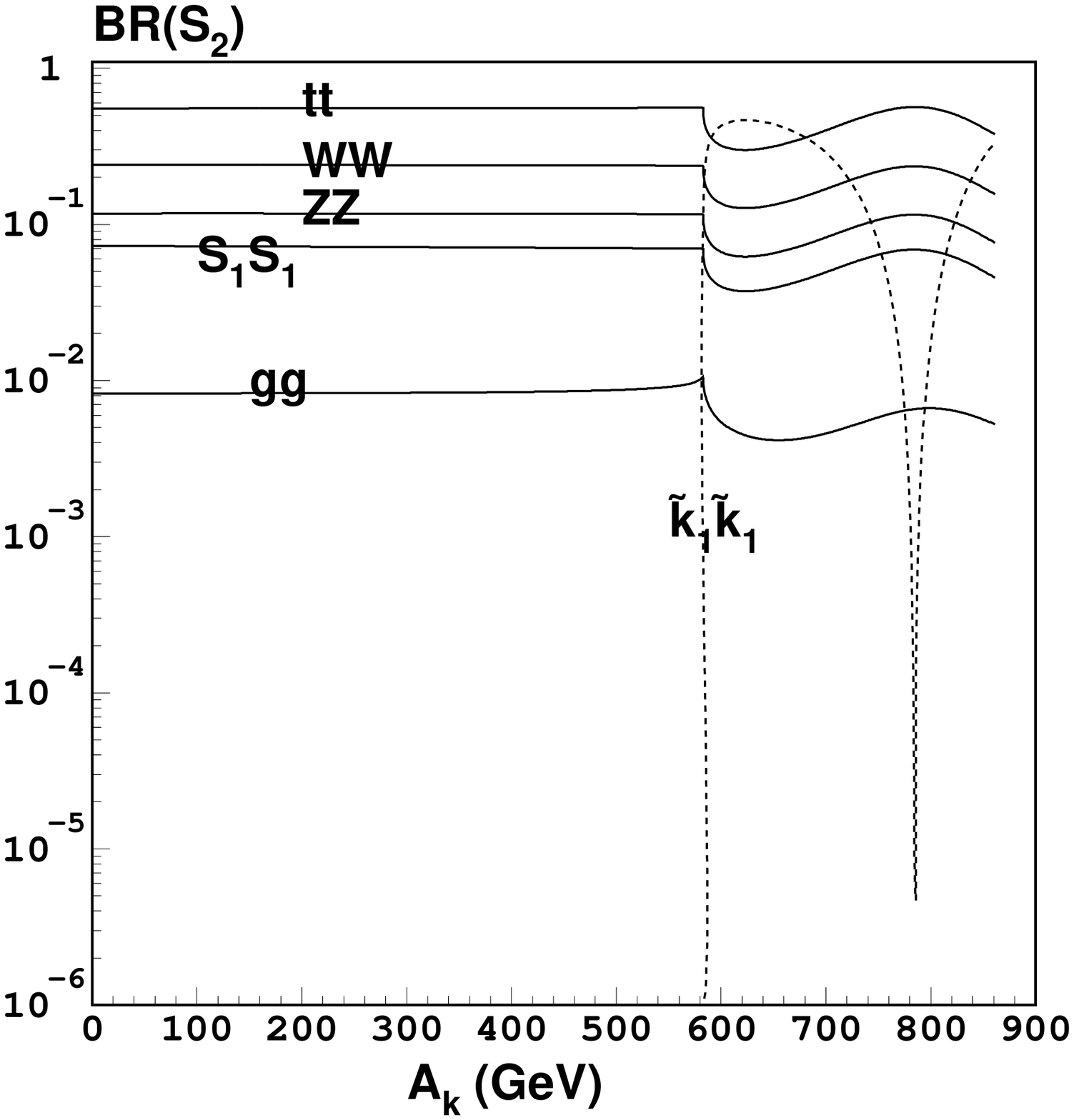}
\caption[plot]{The same as Fig. 5, except that the parameter values are those of Fig. 7}
\end{center}
\end{figure}

\setcounter{figure}{0}
\def\figurename{}{}%
\renewcommand\thefigure{FIG. 11}
\begin{figure}[t]
\begin{center}
\includegraphics[scale=0.6]{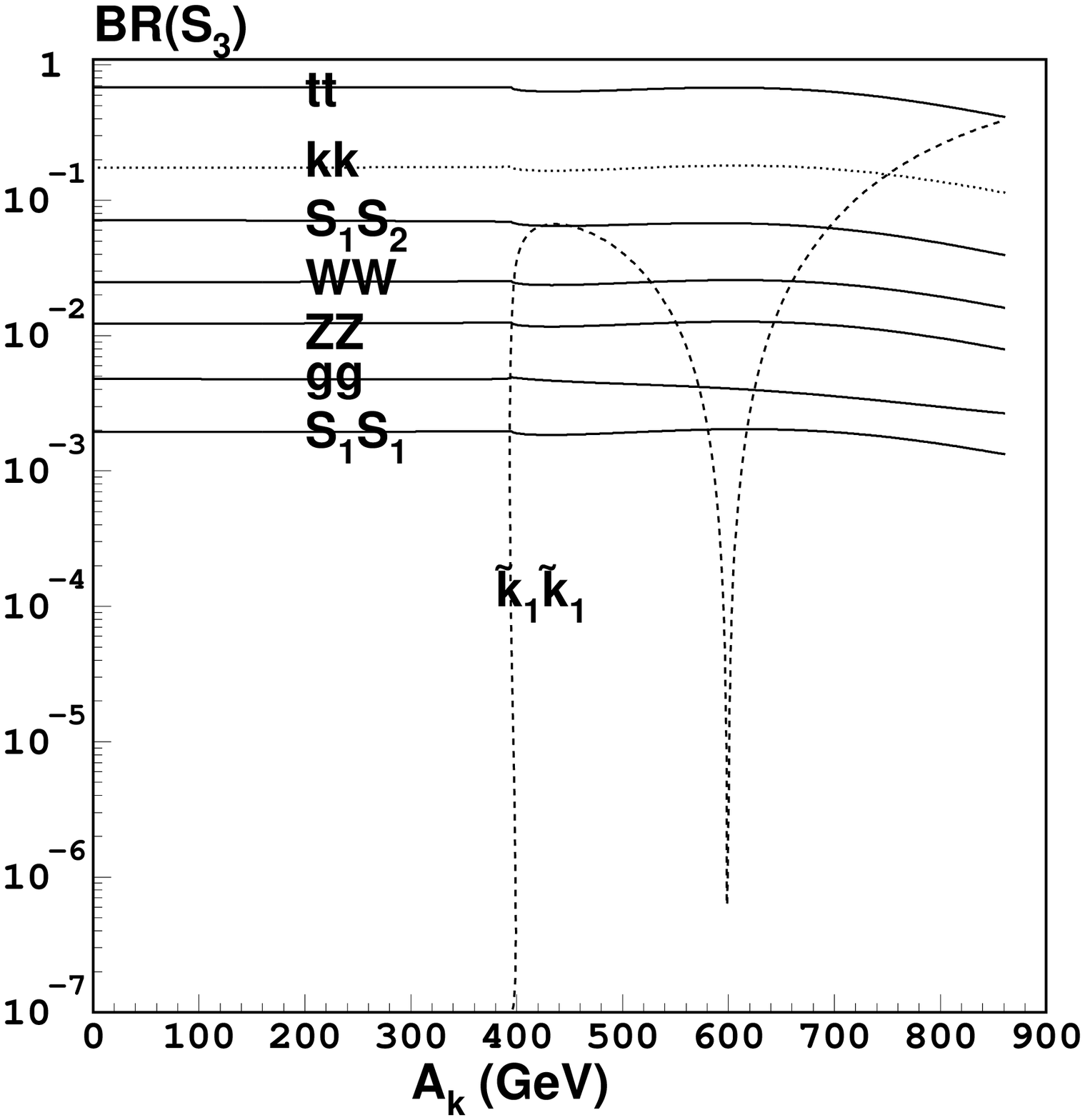}
\caption[plot]{The same as Fig. 6, except that the parameter values are those of Fig. 7.
Here, note that the branching ratio for a pair of exotic quark pairs, BR($S_3 \to k{\bar k})$ is present.
It is the second largest branching ratio, larger than BR($S_3 \to WW$), BR($S_3 \to ZZ$), or BR($S_3 \to gg$).}
\end{center}
\end{figure}

\end{document}